\documentclass[review]{elsarticle}
\usepackage[utf8]{inputenc}
\usepackage{amsmath}
\usepackage{amsfonts}
\usepackage{url}
\usepackage{graphicx}
\usepackage{lineno}
\modulolinenumbers[5]

\usepackage{amssymb}

\begin{document}
\begin{frontmatter}

\title{Artificial Musical Intelligence: A Survey}

\author[a,b]{Elad Liebman
\cortext[mycorrespondingauthor]{Corresponding author}}
\ead{eliebman@sparkcognition.com}

\author[a]{Peter Stone}
\ead{pstone@cs.utexas.edu}

\address[a]{Computer Science Department, The University of Texas at Austin}
\address[b]{SparkCognition Research, Austin, TX}

\label{chap:taxonomy}

\begin{abstract}


Computers have been used to analyze and create music since they were first introduced in the 1950s and 1960s. Beginning in the late 1990s, the rise of the Internet and large scale platforms for music recommendation and retrieval have made music an increasingly prevalent domain of machine learning and artificial intelligence research. While still nascent, several different approaches have been employed to tackle what may broadly be referred to as ``musical intelligence.'' This article provides a definition of musical intelligence, introduces a taxonomy of its constituent components, and surveys the wide range of AI methods that can be, and have been, brought to bear in its pursuit, with a particular emphasis on machine learning methods.


\end{abstract}

\begin{keyword}
Artificial Musical Intelligence;
Music and Artificial Intelligence;
Music Informatics;
\end{keyword}

\end{frontmatter}

\section{Introduction}
\label{intro}

Since its emergence in the 1950s, artificial intelligence has become an ever 
more prevalent field of scientific research. Technology to which we may assign 
varying degrees of intelligence is virtually all around us -- from from 
sophisticated navigation systems\cite{duckham2003simplest} and anti-collision sensors placed on cars \cite{wolterman2008infrastructure} to 
recommender systems meant to help us pick a book or movie\cite{adomavicius2005toward}. However, 
while great emphasis has been placed on improving the performance of such systems, other meaningful facets have not been as thoroughly explored. 
Such additional facets cover a wide array of complex mental tasks 
which humans carry out easily, yet are hard for computers to mimic. These include the human ability to understand social and cultural cues, to 
interpret and infer hidden meanings, to perceive the mental state of their 
counterparts, and to tailor their responses accordingly. A prime example for 
a domain where human intelligence thrives, but machine understanding is 
limited, is music.

%
%
%
%
%
%

In recent years, the application of algorithmic tools in cultural domains 
has become increasingly frequent. An interesting example  
is Sisley the Abstract Painter, a project aimed to algorithmically emulate 
modern paintings of varying abstraction levels, given an input photograph\cite{zhao2010sisley}. 
Another example uses visual recognition tools to study what makes 
the architectural styles in different cities distinctive\cite{doersch2012makes}. A more mainstream 
example for the application of machine learning in a cultural domain 
can be seen in a recent paper in which 16 episodes from the famous 
TV series Lost are automatically tagged for character presence using weakly supervised data \cite{cour2009learning}. In the domain of natural language processing, many 
works have used literary texts as input data, and some works have cultural 
goals such as document style classification \cite{argamon2003gender}, authorship attribution \cite{stamatatos2009survey},
and literature analysis \cite{kirschenbaum2007remaking}. A theme that surfaces from examining this type 
of research is that tools from the AI and machine learning communities often reveal 
new insights and shed new light on fundamental cultural questions -- 
what characterizes an author (or an architect); which geometric properties 
best separate Kandinsky from Pollock (Or Steven Spielberg from Stanley 
Kubrick); is it possible to chart the evolution of Latin dance styles, etc. 
Another important observation is that such cultural domains may often prove 
excellent testbeds for new algorithmic tools and approaches. 

There are many ways in which artificial intelligence and music intersect, ranging from analysis of large bodies of existing music to the creation of music itself. Computers have accompanied both the analysis and creation of music almost since they first came into existence. In 1957, Ljaren Hiller and Leonard Isaacson developed software that applied Markov chains and rule-based logic to compose a string quartet \cite{hiller1959experimental}. Iannis 
Xenakis used computers in the early 1960s to generate numerical patterns, 
which he later transcribed into sheet music \cite{xenakis1992formalized}, and later led the development of the first music programming language, the Stochastic Music Programme (SMP) \cite{xenakis1965free}. A decade later, Pierre Boulez 
founded IRCAM (Institut de Recherche et Coordination Acoustic/Musique), 
where composers, computer scientists, and engineers study music and sound 
and invent new tools for creating and analyzing music \cite{born1995rationalizing}. Only a few years 
after its establishment, IRCAM already served as the home of the Spectralist 
movement, where composers such as Gerard Grisey and Tristan Murail used 
computers and spectral analysis to compose new music \cite{anderson2000provisional}. Since then, the notion 
of applying artificial intelligence to create music has remained of interest 
to many, and there are many other examples for this type of composition, 
ranging from stochastic generation tools and elaborate mathematical models 
to grammar-based generation and evolutionary approaches \cite{jackendoff1972semantic,munoz2016memetic,quick2010generating}.

Another recent body of work lying at the intersection between artificial 
intelligence and music analysis is that of the music information retrieval (or 
MIR) community. Over the last decade, many researchers have applied 
computational tools to carry out tasks such as genre identification \cite{doraisamy2008study}, music 
summarization \cite{mardirossian2006music}, music database querying \cite{eric2003name}, melodic segmentation 
\cite{pearce2008comparison}, harmonic analysis \cite{chen2011music}, and so on. Additional research questions with implications 
for preference learning and computational musicology include (but 
are not limited to) performance analysis and comparison \cite{liebman2012phylogenetic}, music information 
modeling\cite{conklin1995multiple}, music cognition\cite{krumhansl2001cognitive}, and surprise\cite{abdallah2009information}. 

Indeed, the study of music perception within the cognitive science community 
has also served as a bridgehead between computational learning research 
and music analysis. Considerable effort has been put into using algorithmic 
tools to model patterns of psycho-physical responses to music stimuli \cite{juslin2008emotional}, 
and the interaction between musical concepts and their interpretations in 
the brain\cite{krumhansl2001cognitive}. Another related field of study is that of human-computer 
interaction and human-robot interaction. Previous work has been carried 
out in order to provide AI with the ability to interact with humans in one 
form of social setting or another\cite{dautenhahn1995getting}. These works, however, usually do 
not capture the complexity of human interaction, and more importantly, 
rarely take into account the complex array of pre-existing cultural knowledge 
that people ``bring to the table'' when they interact socially, or the cultural 
information they accrue through interaction. 


The separate fields and perspectives on music informatics, spanning music information retrieval, cognitive science, machine learning and musicology, have largely proceeded independently.  However, they are all concerned with overlapping facets of what we define in this survey as ``musical intelligence'', specifically in the context of artificial intelligence. To define something as complex and as abstract as ``musical intelligence'' is at least as difficult as defining what intelligence is in general - a notoriously slippery and tenuous endeavor. However, for the purposes of this article, we adopt the following working definition: 

\begin{quote}
\label{music_intelligence_def}
``Musical Intelligence'', or ``music understanding'', describes a system capable of reasoning end-to-end about music. For this purpose, it needs to be able to reason at different levels of abstraction with respect to music; from perceiving low-level musical properties, to intermediate levels of abstraction involving the organizational structure of the music, to high level abstractions involving theme, intent and emotional content.

\end{quote}

The breakdown of musical abstractions as ``low-level'', ``intermediate'' and ``high-level'' is rather murky. Nonetheless, we can consider basic auditory properties regarding the overall spectrum, tempo, instrumentation etc.\ as the lowest levels  of music understanding. Intermediate levels of abstraction include concepts such as identifying melody vs. accompaniment, identifying the functional harmonic structure of musical segments, identifying recurring motifs, or placing the music in broad genre terms. High-level abstractions include more principled motific and thematic analysis, understanding the intended emotional valence of various pieces of music, the interplay between different structural and motific choices, drawing connections between different pieces of music, recognizing the style of individual musicians, being able to successfully characterize the musical tastes of others, and ultimately, being able to generate musical sequences that people would consider meaningful.

While somewhat analogous to the notion of scene understanding in machine vision \cite{li2009towards}, musical intelligence is a much more elusive term, given that the ``objects'' music deals with are considerably less concrete or functionally defined than those usually considered in computer vision. Nonetheless, the definition above is useful in providing a high-level motivation and goal for connecting disparate aspects of music informatics research.

The purpose of this survey article is threefold. First, it is meant to serve 
as an updated primer for the extremely large and interdisciplinary body of 
work relating to artificial musical intelligence. Second, it introduces 
a detailed taxonomy of music related AI tasks that is meant to provide a 
better perspective on the different achievements made in the intersection of 
both worlds in the past two decades. Third, this survey analyses different 
evaluation methods for various music-related AI tasks.

Due to the enormous literature that is relevant to this survey, we limit its scope in several ways. We focus on works that involve a significant machine learning or artificial intelligence component. We focus on several potential representations of music, either symbolic or at the audio level, and consider tasks primarily involving this input. While we acknowledge the large body of work which focuses on the signal-processing and audio-extractive aspects of automated music analysis, we do not consider it a core part of this survey, and only reference it to the extent that such work lies at the heart of many of the feature extraction procedures used in machine learning frameworks for music related tasks. Another large body of work, which focuses on natural language processing of song lyrics, music reviews, user-associated tags etc.\ is also considered outside the scope of this article. We also consider automated sheet music recognition (traditionally through image processing techniques) as outside the scope of this survey.

The structure of this article is as follows: in Section \ref{chap8:back} we discuss the target 
audience of this survey, and provide initial background for reading the article. We proceed to discuss the motivation behind music-related AI research and its potential uses.
In Section \ref{chap8:tax} we focus on the proposed taxonomy and break down the extensive body of literature into different categories,
utilizing different perspectives. Subsequently, in Section \ref{chap8:tasks} we review the literature from the perspective of the tasks that have been studied. In Section \ref{chap8:repr} we discuss the different types of representations that have been used in the literature. In Section \ref{chap8:technique} we break down the extensive list of AI techqniques that have been applied in music AI research.
Section \ref{chap8:eval} discusses the different evaluation methods used in 
the literature to assess the effectiveness of proposed approaches. Lastly, in Section \ref{chap8:musint} we summarize the contributions of this survey, consider the idea of artificial musical intelligence from a broader perspective, and discuss potential gaps existing in the literature.

\section{Background and Motivation}
\label{chap8:back}
This survey article is 
aimed at computer scientists working in AI who are interested in music as a 
potential research domain. Since both the study of music and the artificial intelligence and machine learning 
literature are too extensive to be covered by any single survey paper, we assume the reader has at least some knowledge about the basic machine learning paradigm (e.g. supervised vs. unsupervised learning, partition to training and testing data, evaluative metrics for learning algorithms such as area under the ROC curve etc). We also assume some familiarity with various learning architectures and algorithms, such as regression, support vector machines, decision trees, artificial neural networks, probabilistic and generative models, etc. From a more classical AI viewpoint, some understanding of knowledge representation, search and planning approaches is assumed, but is not directly relevant to a large component of this paper. Good introductory sources for machine learning and AI concepts can be found in various textbooks (such as \cite{russell1995modern}).

Regarding music terminology, we assume familiarity with a few basic terms. These terms include pitch, note, scale, key, tempo, beat, chord, harmony, cadenzas and dynamics. We provide brief explanations for these terms and more in Appendix \ref{app:glossary}. Further details can be found in sources such as The Oxford dictionary of musical terms \cite{latham2004oxford}, among many others. Throughout the article we will assume the general meaning of these terms is known.

%



This survey lays down the case that work at the intersection of artificial intelligence and music understanding is beneficial to both communities on multiple levels. As a rich, complex research domain, we expect that the study of artificial musical intelligence will potentially produce fundamental scientific discoveries, as well as engineering insights and advances which could be applicable in other domains. These expectations are supported by the following lines of reasoning:


\begin{itemize}
\item {\bf Music is a quintessential form of intelligence:} Music, being intrinsically complex and multifaceted, involves some of the most sophisticated mental faculties humans have. Musical skills such as playing, analyzing or composing music involve advanced data analysis, knowledge representation and problem solving skills. The challenge of developing such skills in artificial agents gives rise to interesting research problem, many of which are transferable to other application domains (such as analyzing video or interactive gameplay). Furthermore, some abstract issues such as modeling a ``utility function'' that captures a person or a group's enjoyment of different types of musical information are in fact inherent to any attempt to quantify aesthetic value, mass appeal or creative content. Advances in the modeling of such a function would have immediate applications in any case where understanding ``what people want'' is key to good performance but no easily quantifiable objective functions exist.

\item {\bf Music is inherent to the human experience, and therefore to social interaction:} If we envision a future where intelligent artificial agents interact with humans, we would like to make this interaction as natural as possible. We would therefore like to give AI the ability to understand and communicate within cultural settings. This issue has actual benefits, as software capable of tailoring its behavior to the tastes and the preference of specific people would do better both in understanding the behavior of its human counterpart and influence it, leading to a much more successful interaction.

\item {\bf Deeper music AI will lead to better performance of real world systems:} Let us consider a recommendation system for music. Such a system would greatly benefit from the ability to model the intrinsic properties of the music it deals with, rather than solely rely on statistical correlations or simplistic measures.
This capacity would also enable recommendation models to learn with less input data, thus ameliorating the infamous cold start problem in recommender systems. The work of Liebman et al. \cite{DJMC} is an example for this approach. The architecture presented in that work is able to learn some basic signal of what a person likes based on very little experience by directly mapping musical properties of songs and transitions to predicted human preferences.

\item {\bf AI can lead to new cultural insights:} The intersection of artificial intelligence and music often leads to insights regarding music, how it is perceived by humans, and what makes it unique. These observations have significant cultural value, and are of interest to many researchers in a wide range of disciplines.

\end{itemize}

While admittedly the study of musical intelligence can be seen as somewhat more esoteric than other core academic disciplines and application areas, and the assessment of musical quality is inherently subjective, to those concerned about such issues we offer the following observations:


\begin{itemize}
\item {\bf Widespread commercial interest:} The market for music recommendation, for instance, is large\footnote{\url{http://techcrunch.com/2015/01/21/apple-musicmetric}}, and growing. Video games such as Rocksmith\footnote{\url{http://en.wikipedia.org/wiki/Rocksmith}} , which automatically analyzes actual instrument playing to provide feedback and accompaniment, are also growing in popularity. The commercial success of such applications reflects a strong industrial interest in research that enables better autonomous music understanding.

\item {\bf Widespread academic interest:} In the past two decades, there have been hundreds of papers at the intersection of AI and music published in top tier conferences and journals (including those which we discuss in this survey), with thousands of citations, cumulatively. This fact in itself serves as evidence for the existing interest for such work across varied research communities.

\item {\bf Realizable goals exist:} While the subjectivity inherent to music may pose difficulties in evaluating the performance of various music AI systems, many inter-subjective goals (such as increasing user satisfaction and engagement, or better matching people's perceptions and expectations) can be effectively evaluated using lab experiments and crowd-sourcing.
\end{itemize}

\section{A Taxonomy of Music AI Problems}
\label{chap8:tax}

Consider a song by the Beatles, or a piano trio by Beethoven. What kinds of computational research questions can we ask about these cultural artifacts? What kinds of tasks might we expect intelligent software to perform with respect to them? 


Due to the complexity and richness of the music domain, countless different 
perspectives can be assumed when studying the intersection of music and artificial intelligence. Different perspectives give rise to different research questions and different approaches. In order to compare and contrast the literature using a consistent and unified language, we introduce the following dimensions along which each contribution can be placed: 
\begin{itemize}
\item The target task
\item The input type
\item The algorithmic technique(s)
\end{itemize}

\begin{figure}[!htb]
\centering
\includegraphics[width=.8\linewidth]{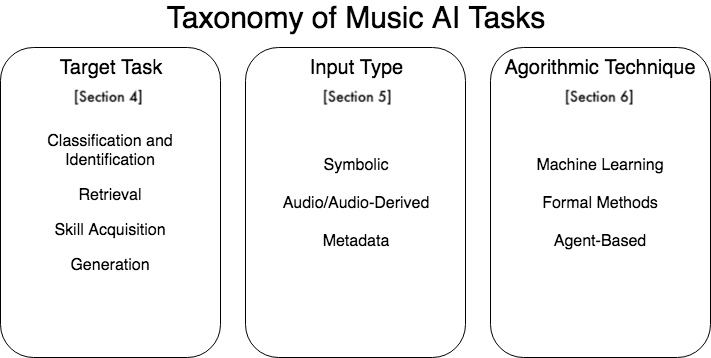}
\caption{Visual high-level illustration of the proposed taxonomy.} \label{chap8:fig_tax}
\end{figure}

In this section we broadly outline these three perspectives, which together span the taxonomy introduced in this survey. A visual representation of the proposed taxonomy is shown in Figure \ref{chap8:fig_tax}.

	\subsection{Partition by the Nature of the Task}
	
There is a wide variety of potential research tasks we might concretely try to accomplish in the music domain. We use the term ``task'' to describe a small, concrete and well-defined problem. For instance, in the Beatles song example above, we may wish to discern the chorus from the refrain, or identify the beat and the key of the song, or identify whether it is an early vs. a late Beatles song. While these are all small and concrete tasks, they are not atomic or independent; knowing the key and the beat of a song is relevant both to determining its structure (chorus vs. refrain), identifying which sub-genre it belongs to, etc.

To better understand shared themes across tasks and facilitate a more helpful overview of the current literature, we break tasks down to several categories:
\begin{enumerate}

\item Classification and Identification - any tasks which associate musical segments with one or more out of a closed set of labels. For example, classifying pieces by composer and/or genre. 

\item Retrieval - as in the broader case of information retrieval, these tasks involve obtaining relevant items, often ranked by relevance, from a music dataset. A typical example is a recommender system that suggests specific songs to a specific user given his listening history.

\item Musical Skill Acquisition - this category encompasses the varied set of basic analysis skills required for music processing, from pitch and tempo extraction to chord recognition.

\item Generation - these tasks involve some facet of creating new musical expression, ranging from generating expressive performance from audio, generating meaningful playlists by sequencing existing songs, and, probably the most elusive of all, generating new music.

\end{enumerate}

These categories aren't mutually exclusive, as many actual tasks might share more than one aspect, or contain components that belong in other categories. Still, we believe it is a natural way to group tasks in a way that sheds light on recurring themes and ideas.

	\subsection{Partition by Input Type}

It is almost always the case that the type of input dramatically affects the range and complexity of tasks which can be performed on that input. Generally, there are three input categories --

\begin{enumerate}
\item Symbolic Music Representations - these are the simplest and easiest 
to analyze, as they capture pitched event information over time. Variants of symbolic representation range from the ubiquitous MIDI protocol \cite{loy1985musicians} to complex digital representation of common practice music notation.

\item Audio (and audio-derived) Representations -  this category of representations ranges from raw unprocessed audio to compressed audio to concise spectral features, depending on the level of reduction and abstraction.

\item Related Musical Metadata - all non-audio information regarding a musical piece, ranging from artist and song names to associated mood tags, social media information, lyrics, occurrence history etc.
\end{enumerate}

In this survey we will focus on the first two representations, but 
due to its ubiquity, we will occasionally refer to the third type.

	\subsection{Partition by Algorithmic Technique}

A wide variety of machine learning and artificial intelligence paradigms and techniques have been applied in the context of music domains.	 From a machine learning and artificial intelligence research perspective, it is of interest then to examine this range of techniques and the specific musical domains where they were applied. Due to the extensive nature of the related literature and the wide range of musical tasks where the following methods have been used, this list is not intended to be entirely comprehensive. To the best of our knowledge, however, it is representative of the full array of methods employed. Broadly speaking, we consider the following general technical approaches:

\begin{enumerate}

\item Machine Learning Approaches - a wide range of machine learning paradigms has been employed for various music informatics tasks. The list of techniques used is as varied as the machine learning literature itself, but some examples include methods such as support vector machines (SVM) \cite{hearst1998support}, generative statistical models such as Hidden Markov Models (HMM) \cite{rabiner1989tutorial}, Markov Logic Networks (MLN) \cite{richardson2006markov}, Conditional Random Fields (CRF) \cite{lafferty2001conditional}, and Latent Dirichlet Allocation (LDA) \cite{blei2003latent}. In recent years, deep neural network architectures such as Convolutional Neural Networks (CNN) \cite{lecun1995convolutional}, Recurrent Neural Networks (RNN) \cite{gurney1997introduction}, and Long Short Term Memory networks (LSTMs) \cite{hochreiter1997long} have become increasingly popular and ubiquitous in the music informatics literature. 

\item Formal methods - multiple attempts have been made to employ formal techniques, similar to the formal methods subfield in computer science, to handle music informatics tasks via formal specification methods to describe and generate musical sequences. Under this umbrella one may find approaches inspired by generative grammars \cite{jackendoff1972semantic}, formal specification of tonal and chordal spaces with production rules \cite{davis1977production}, probabilistic logic \cite{von1952probabilistic}, and fuzzy logic \cite{zadeh1975fuzzy}.

\item Agent-based techniques - multiple papers in the music AI literature have studied complex approaches that go beyond the scope of a learning algorithm or a formal specification, but rather fall in the subfield of intelligent agent research. That is to say, this category deals with AI systems that combine perception and decision-making in a nontrivial manner to handle musical tasks. In this category one may find examples such as person-agent accompaniment and improvisation systems \cite{thom2000bob}, robotic systems for music performance \cite{shimon}, multiagent music generation architectures \cite{blackwell2003swarm}, and reinforcement learning agents for music generation \cite{cont2006anticipatory}.

\end{enumerate}


Having outlined the general structure of the taxonomy proposed in this survey, we can now delve more deeply into each category and provide examples for the varied types of questions and approaches studied in the past 15 years, following the rise of online music platforms and medium-to-large-scale music datasets. In the next sections we consider each dimension of the taxonomy separately and overview key examples in each partition category.

	\section{Overview of Musical Tasks}
\label{chap8:tasks}
\begin{figure}[!htb]
\centering
\includegraphics[width=.8\linewidth]{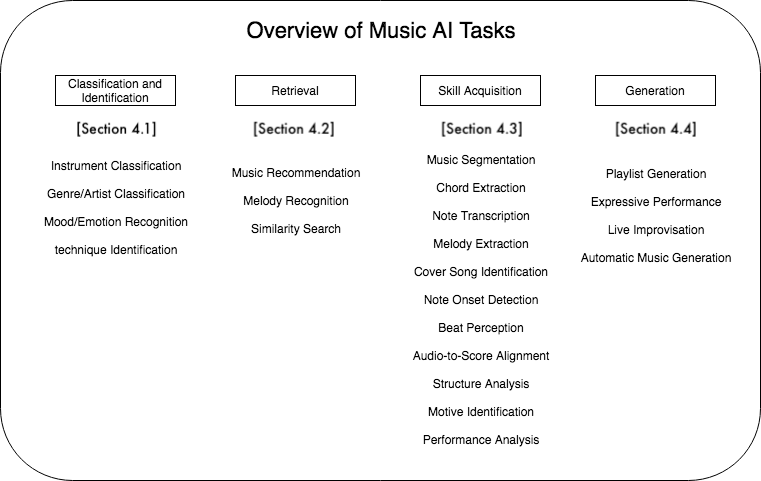}
\caption{Visual high-level illustration of music AI tasks.} \label{chap8:fig_tasks}
\end{figure}

The first aspect through which we examine the literature is the functional one - which musical tasks have been addressed via machine learning and artificial intelligence approaches? Following our taxonomy from Section \ref{chap8:tax}, we partition tasks into four main groups - classification and identification, retrieval, skill acquisition, and generation. A visual summary of the content surveyed in this section is provided in Figure \ref{chap8:fig_tasks}.

		\subsection{Classification and Identification Tasks}
\label{chap8:classification}
Suppose we are presented with a newly unearthed score for a classical piece. This score, it is claimed, is a lost cantata by Johann Sebastian Bach, one of the many assumed to have been lost to posterity. Is this really lost music by the great Baroque master? Or perhaps the work of a talented imposter? Was it written in Bach's time? Or is it a recent work of forgery? These questions may seem hypothetical, but they are actually quite real, for instance in the case of several organ fugues by J.S. Bach \cite{van2008assessing}. An even more famous example involving J.S. Bach, one that most music students should be familiar with, is that of Bach's famous liturgical chorales. Of these 327 chorales, which have been widely used to teach traditional harmony and counterpoint for more than two centuries, only a third have definite sources in known Bach cantatas. The others are without a known source, and many have speculated that at least some of them were written by Bach's students (indeed, many have disputed the authorship of entire Bach cantatas, for instance \cite{owen1960authorship}). If we had a reliable way to computationally predict the likelihood that a previously unknown piece was actually written by Bach (vs., say, any other of Bach's many contemporaries), it would help greatly not only in shedding light on such musicological mysteries, but also in revealing what it is that makes Bach unique.

Music domains offer a wide array of potential \emph{classification tasks}. Therefore, partly due to their ease of evaluation (as we discuss further in Section \ref{chap8:eval}), they have been a mainstay of the music informatics field for several decades. Indeed, surveying the literature from the past 15 years, a varied list of classification tasks emerges.

Early examples for modern approaches include Scheirer and Slaney, who compared various machine learning techniques, including maximum-aposteriori (MAP) estimators, Gaussian Mixture Models, feature space partitioning and nearest-neighbor search, in order to discriminate speech from music based on acoustic features \cite{scheirer1997construction}. Another such early example is the work of Marques and Moreno, who tackled the issue of instrument classification using Gaussian mixture models and SVMs \cite{marques1999study}. 

\emph{Instrument classification} has been a common thread in the music information retrieval literature. In a survey from 2000, Herrera et al \cite{herrera2000towards} point out several machine learning techniques already employed to identify which instrument is playing in solo recordings. Examples of such works include K-nearest neighbors (KNN), employed for example by Martin and Kim \cite{Martin98musicalinstrument}, Naive Bayes classifiers (see Martin \cite{martin1999sound}), and support vector machines (SVMs) (see Marques \cite{marques1999automatic}). Eichner et al.\ have introduced the usage of Hidden Markov Models for this purpose in a more realistic and diversified setting with multiple instruments of the same kind \cite{eichner2006instrument}. In their experiments, they inferred HMMs in which the states are Gaussian probability density functions for each individual instrument and for each individual note, comparatively, in a data-driven manner, and were able to show that for their particular dataset of real-world recordings, this approach outperformed the baselines. Benetos et al.~\cite{benetos2006musical} applied Nonnegative matrix factorization and subset selection, resulting in improved classification accuracy compared to results obtained without these modifications. Joder et al.~\cite{joder2009temporal} introduced the notion of {\it temporal integration} to instrument recognition. Simply put, temporal integration involves the combination of features across multiple time frames to construct more context-aware, higher-level features (the notion was first introduced in a music domain by Meng et al.~\cite{meng2007temporal}). By combining temporally aggregated features they were able to beat the state of the art for that time. Considering the harder problem of multi-instrument classification, Garcia et al.\ were able to classify multiple instruments as long as they were in separate recording channels (with some tolerance to leaking) by training statistical models for individual partials per instrument class \cite{garcia2011simple}. In more recent work, Fourer et al.~\cite{fourer2014automatic} took a hierarchical approach to classifying timbre in ethnomusicological audio recordings. Their method introduces a hierarchical taxonomy from instruments to sound production categories, which bifurcate further (aerophones $\rightarrow$ blowing; cordophones $\rightarrow$ bowing; plucking or striking; etc), and embeds each timbre class in a projection space that captures weights over these descriptors (the training is done via Latent Discriminant Analysis \cite{mika1999fisher}). 


The issue of instrument classification ties in organically to another prevalent line of research, that of \emph{genre classification}. Tzanetakis et al.\ introduced a hierarchy of audio classification to speech vs. music, genres, and subgenres \cite{george2001automatic}. Using timbral audio features, they were able to reach accuracy of ~60\% using Gaussian mixture models. Dubnov et al.~\cite{dubnov2003using} trained statistical models to describe musical styles in a way that could also be harnessed towards music generation (a topic we expand on in subsection \ref{chap8:gen}). Their approach employs dictionary-based prediction methods to estimate the likelihood of future sequences based on the current context (in the paper they compare incremental parsing to the more sophisticated predictive suffix trees). In a comparative study from the same year as the Dubnov et al.\ work, Li et al.\ compared multiple audio feature sets and classifier variations (based on SVMs, KNN, GMM and discriminant analysis), and across several different datasets \cite{li2003comparative}. In 2007, Meng et al.\ studied the application of temporal integration (a method we mentioned in the paragraph above) to genre classification \cite{meng2007temporal}, leading to improvements in performance and robustness. 

Different researchers have taken different perspectives on the issue of \emph{finding useful representations} for genre classification (an issue we also discuss in Section \ref{chap8:repr}). For instance, Panagakis et al.\ applied nonnegative multilinear PCA to construct better features for genre classification \cite{panagakis2010non}, while Salamon et al.\ used melody extraction for genre classification in polyphonic settings, reaching accuracy of above 90\% on a small sample of distinct genres (instrumental Jazz, vocal jazz, opera, pop, and flamenco) \cite{salamon2012musical}. Anan et al.\ used a theoretically grounded approach for learning similarity functions for the purpose of genre recognition \cite{anan2012polyphonic}. In order to train these similarity functions, they converted their MIDI input to binary chroma sequences (a single binary chroma vector is a sequence of length 12 for each tone, in which present tone indices are assigned the value of $1$ and the rest are $0$). Marques et al.\ applied optimum path forests (OPF), a graph-partitioning ensemble approach, for genre classification in potentially large and dynamic music datasets \cite{marques2011new}. Rump et al.\ separated harmonic and percussive features in recordings with autoregressive spectral features and SVMs to improve performance over a non-separated baseline \cite{rump2010autoregressive}, while Panagakis et al.\ used locality-preserving nonnegative tensor factorization as another means of constructing better features for genre classification \cite{Panagakis2010sparse}. In contrast, West and Cox studied the issue of optimizing frame segmentation for genre classification \cite{west2004features} (we discuss the issue of segmentation more in-depth in Section \ref{chap8:skill_acq}). Arjannikov et al.\ tackled the issue of genre classification from a different perspective by training an associative classifier \cite{arjannikov2014association} (conversely, association analysis in this context can be perceived as KNN in multiple learned similarity spaces). Hillewaere et al.\ applied string methods for genre classification in multiple dance recordings, transforming the melodic input into a symbolic contour sequence and applying string methods such as sequence alignment and compression-based distance for classification \cite{hillewaere2012string}. Somewhat apart from these works, Mayer and Rauber combine ensembles of not only audio but also lyric (i.e.\ textual) features for genre classification \cite{mayer2011musical}. In more recent work, Herlands et al.\ tackled the tricky issue of homogenous genre classification (separating works by relatively similar composers such as Haydn and Mozart), reaching accuracy of 80\% using specifically tailored melodic and polyphonic features generated from a MIDI representation \cite{herlands2014machine}. Interestingly, Hamel et al.\ also studied the issue of transfer learning in genre classification, demonstrating how classifiers learned from one dataset can be leveraged to train a genre classifier for a very different dataset \cite{hamel2013transfer}.







Another common classification task in the music domain is that of \emph{mood and emotion recognition} in music, a task which is interesting both practically for the purpose of content recommendation, and from a musicological perspective. Yang and Lee used decision trees to mine emotional categorizations (using the Tellegen-Watson-Clark mood model \cite{tellegen1999dimensional}) for music, based on lyrics and tags, and then applied support vector machines to predict the correspondence of audio features to these categories \cite{yang2004disambiguating}. Han et al.\ applied support vector regression to categorize songs based on Thayer's model of mood \cite{thayer1990biopsychology}, placing songs on the Thayer arousal-valence scale \cite{han2009smers}. Trohidis et al.\ also used both the Tellegen-Watson-Clark model and Thayer's model, and reframed the emotion classification problem as that of multilabel prediction, treating emotional tags as labels \cite{trohidis2008multi}. Lu et al.\ applied boosting for multi-modal music emotion recognition \cite{lu2010boosting}. In their work, they combined both MIDI, audio and lyric features to obtain a multi-modal representation, and used SVMs as the weak learners in the boosting process. Mann et al.\ classified television theme songs on a 6-dimensional emotion space (dramatic-calm, masculine-feminine, playful-serious, happy-sad, light-heavy, relaxing-exciting) using crowd-sourced complementary information for labels, reaching accuracy of 80-94\% depending on the emotional dimension \cite{mann2011music}. Focusing on audio information alone, Song et al.\ studied how different auditory features contribute to emotion prediction from tags extracted from last.fm \cite{song2012evaluation}. Recently, Delbouys et al. proposed a bimodal deep neural architecture for music mood detection based on both audio and lyrics information \cite{delbouys2018music}.

It is worth noting that obtaining ground truth for a concept as elusive as mood and emotion recognition is tricky, but labels are often obtained through mining social media or through crowdsourcing, under the assumption that people are the ultimate arbiters of what mood and emotion music evokes. We discuss this matter in greater detail in section \ref{chap8:eval}.



The works described above are a representative, but not comprehensive, sample of the type of work on music classification that has taken place in the last 15 years. Various other classification tasks have been studied. To name a few, Su et al.\ recently applied sparse cepstral and phase codes to identify guitar playing technique in electric guitar recordings \cite{su2014sparse}; Toiviainen and Eerola used autocorrelation and discriminant functions for a classification based approach to meter extraction \cite{toiviainen2005classification}; several works including that of Lagrange et al.\ tackled the issue of singer identification \cite{lagrange2012robust}, while Abdoli applied a fuzzy logic approach to classify modes in traditional Iranian music recordings \cite{abdoli2011iranian}.


		\subsection{Retrieval Tasks}

Consider now that you are in charge of picking music for a specific person. The only guidance you have is that previously, that person listed some of the songs and the artists he likes. Provided with this knowledge, your task is to find additional music that he will enjoy. You can accomplish this goal by finding music that is \emph{similar} to the music he listed, but different. For this purpose, you must also define what makes different pieces of music similar to one another. Alternatively, you may be faced with a recognition task not that far removed from the classification tasks we listed in the previous subsection: given a piece of music, find a subset of other musical pieces from a given corpus which are most likely to have originated from the same artist. These are just a couple of examples for music retrieval tasks, which combine music databases, queries, and lower-level understanding of how music is structured.

In this subsection we consider different types of retrieval tasks in musical context. These tasks usually require a system to provide examples from a large set given a certain query. Selecting examples that best suit the query is the main challenge in this type of task.

The most straightforward context for such retrieval tasks is that of \emph{music recommendation}. Given some context, the system is expected to suggest songs from a set best suited for the listener. This type of task has been a widely studied problem at the intersection of music and AI. Yoshii et al.\ combined collaborative and content-based probabilistic models to predict latent listener preferences \cite{yoshii2006hybrid, yoshii2007improving}. Their key insights were that collaborative filtering recommendation could be improved, first by combining user ratings with structural information about the music (based on acoustic data), revealing latent preferences models; and secondly, by introducing an incremental training scheme, thus improving scalability. Similarly, Tiemann et al.\ also combined social and content-based recommenders to obtain more robust hybrid system \cite{tiemann2007ensemble}. Their approach is ensemble-based, with separate classifiers trained for social data and for music similarity later combined via a learned decision rule. 

A different thread in the music recommendation literature explores the aspect of associating tags with songs. Roughly speaking, tags are a broad set of user-defined labels describing properties of the music, ranging from genre description (``indie'', ``pop'', ``classic rock'' and so forth), to mood description (``uplifting'', ``sad'' etc), to auditory properties (``female vocalist'', ``guitar solo'' etc), and so forth. Along these lines, Eck et al.\ trained boosting classifiers to automatically associate unobserved tags to songs for the purpose of improving music recommendation \cite{eck2007autotagging}. Similarly, Horsburgh et al.\ learned artificial ``pseudo-tags'' in latent spaces to augment recommendation in sparsely annotated datasets \cite{horsburgh2015learning}. More recently, Pons et al. compared raw waveform (unprocessed audio) vs. domain-knowledge based inputs with variable dataset sizes for end to end deep learning of audio tags at a large scale \cite{pons2017end}.

From a temporal perspective, Hu and Ogihara tracked listener behavior over time to generate better models of listener song preferences \cite{hu2011nextone}. Specifically, they use time-series analysis to see how different aspects of listener preference (genre, production year, novelty, playing frequency etc) are trending in order to shape the recommendation weighting. In a related paper, Hu et al.\ also comparatively evaluated how different features contribute to favorite song selection over time \cite{hu2013evaluation}. From the somewhat related perspective of balancing novelty with listener familiarity and preferences, Xing et al.\ enhanced a standard collaborative filtering approach by introducing notions from the multi-armed bandits literature, in order to balance exploration and exploitation in the process of song recommendation, utilizing a Bayesian approach and Gibbs Sampling for arm utility inference \cite{xing2014enhancing}.

A full discussion of the components and intricacies of music recommender systems is beyond the scope of this paper, but can be found in Schedl et al.~\cite{knees2013survey} and Song et al.~\cite{song2012survey}.







Another example for a common retrieval task is that of \emph{melody recognition}, either from examples or via a query-by-humming system. Betser et al.\ introduced a sinusoidal-modeling-based fingerprinting system and used it to identify jingles in radio recordings \cite{betser2007audio}. Skalak et al.\ applied vantage point trees to speed up search of sung queries against a large music database \cite{skalak2008speeding}. A vantage point tree partitions a metric space hierarchically into intersection spheres. By embedding songs in a metric space and using vantage point trees querying can be significantly reduced. Miotto and Orio applied a chroma indexing scheme and statistical modeling to identify music snippets against a database \cite{miotto2008music}. Similar to the representation discussed in Anan et al.~\cite{anan2012polyphonic}, a chroma index is a length $12$ vector which assigns weights for each pitch class based on the Fourier transform of a music fragment. A statistical model representing their chroma frequencies over time is then used with an HMM model for song identification. Another paper that considers identification in a time-series context, but from a different perspective, is that of Wang et al., who iteratively segmented live concert recordings to sections and identify each song separately to recover complete set lists \cite{wang2014automatic}. Also in the context of considering structural properties of music over time, Grosche et al.\ recovered structure fingerprints, which capture longer structural properties of the music compared to standard fingerprints, to improve the retrieval of matching songs from a database given a query \cite{grosche2012structure}. These similarity fingerprints are constructed via self-similarity matrices \cite{foote1999visualizing} on CENS features \cite{muller2005audio}. Recently, Bellet et al.\ introduced a theoretically grounded learned discriminative tree edit similarity model to identify songs based on samples using information about the music semantics \cite{bellet2016learning}. 


The previously mentioned tasks of music recommendation and melody recognition are strongly connected to the key notion of \emph{similarity in music information retrieval}. Given a query, instead of being asked to retrieve the exact same songs, the system may be expected to retrieve songs which are similar to the query. This sort of problem leads to an extensive branch of research on similarity search in music. Platt considered sparse multidimensional scaling of large music similarity graphs to recover latent similarity spaces \cite{platt2004fast}. Similarly inspired, Slaney et al.\ studied various metric learning approaches for music similarity learning \cite{slaney2008learning}, while McFee and Lanckriet proposed a heterogeneous embedding model for social, acoustic and semantic features to recover latent similarities \cite{mcfee2009heterogeneous}. McFee et al.\ also employed collaborative filtering for this purpose \cite{mcfee2010learning}. In a later paper, McFee and Lanckriet expanded the scale of their similarity search approach using spatial trees \cite{mcfee2011large}. Similarly to Mcfee et al., Stenzel and Kamps were able to show that employing collaborative filtering can generate more robust content based similarity measures \cite{stenzel2005improving}. From an entirely different perspective, Hofmann-Engl proposed a cognitive model of music similarity to tackle the complicated and multi-dimensional issue of how we define two pieces of music to be similar, applying general melotonic (pitch distribution) transformations \cite{hofmann2001towards}. Flexer et al.\ studied the modeling of spectral similarity in order to improve novelty detection in music recommendation \cite{flexer2005novelty}. Mueller and Clausen studied transposition invariant self similarity matrices (which we mentioned in the context of Grosche et al.~\cite{grosche2012structure}) for music similarity in general \cite{muller2007transposition}. Hoffman et al.\ studied the application of hierarchical Dirichlet processes to recover latent similarity measures \cite{hoffman2008content}. In that work, each song is represented as a mixture model of multivariate Gaussians, similar to a Gaussian Mixture Models (GMM). However, unlike GMMs, in the Hierarchical Dirichlet Process, the number of mixture components is not predefined but determined as part of the posterior inference process. The hierarchical aspect is derived from the fact that each song is defined by a group of MFCC features. Similarity between songs can be defined according to the similarity between their corresponding distributions over components. In a somewhat conceptually related paper, Schnitzer et al.\ employed ensembles of multivariate Gaussians and self organizing maps to learn a similarity metric for music based on audio features \cite{schnitzer2010islands}. Wang et al.\ used bag of frame representations to compare whole pieces to one another \cite{wang2011learning}. Other approaches include that of Ahonen et al.\ who used a compression based metric for music similarity in symbolic polyphonic music \cite{ahonen2011compression}, and that of Garcia-Diez et al., who learned a harmonic structure graph kernel model for similarity search \cite{garcia2011simple}. In that specific work, binary chroma vectors (dubbed ``chromagrams'' in this paper) are transformed to tonal centroid vectors to reduce the chromagram space from $2^{12}$ to $2^6$. Subsequently, the similarity between query and dataset inputs is measured via the Normalized Compression Distance (NCD) \cite{cebrian2007normalized}. For a specific review of audio-based methods for music classification (specifically, genre and mood classification, artist identification and instrument recognition) and annotation (or auto-tagging, to be more exact), see \cite{fu2011survey}.




		\subsection{Musical Skill Acquisition Tasks}
\label{chap8:skill_acq}
The tasks we described above tend to rely on the ability to effectively represent music in a meaningful way which reflects its property and structure. Such a representation is often obtained through manually designed features (see\cite{berenzweig2001locating} for example). However, a large and varied body of work focuses on the ability to automate the construction of such representations. We consider the spectrum of tasks that lead to useful representations of musical features and structure as \emph{musical skill acquisition}. In the music recommendation example we discussed in the previous subsection, we raised the question of what makes two pieces of music similar to one another, and what makes them distinct. Similarity can lie in basic things like tempo and amplitude, and the overall spectral signature of the piece (what frequencies are heard most of the time). It can lie in subtler things, like how the spectrum changes over time. It can also lie in more abstract musicological properties, such as the rhythmic, harmonic and melodic patterns the music exhibits. Capturing such higher level musical properties is the common thread tying the different tasks we consider as musical skill acquisition tasks.

While the separation between classification or retrieval tasks and ``musical skill acquisition'' is somewhat nuanced, the key distinction is the following. Classification and retrieval tasks reduce music-related problems to a ``simple'' computational question that can be studied directly with its musical aspect abstracted away, as the problem has been reframed as a pure classification or information retrieval problem. On the other hand, in the case of musical skill acquisition, we are interested in training a system to learn some fundamental nontrivial property that involves music. Such a task can be in service of a classification or retrieval task further down the line (for instance, identifying harmonic structure for similarity search) or rely on a lower level classification or retrieval task (for instance, harmonic progression analysis by first classifying individual pitches in each frame), but learning the musical property is in itself the goal and therefore the nature of these tasks is different.
		
Ever since the 18th century, Western scholars have studied the different structural and auditory patterns and properties that characterize music, in what eventually became the modern field of musicology \cite{tomlinson2012musicology}. Musicologists study the structure of melody (how sequences of pitches are combined over time), harmony (how multiple pitches are combined simultaneously over time), rhythm and dynamics. Since the 1960s, musicologists have been using computers to aid in their analyses, when studying large corpora of music or previously unfamiliar music \cite{bel1993computational}. and when focusing on aspects of music that were previously harder to study quantitatively, such as nuances in vibrato or articulation for violin performers \cite{liebman2012phylogenetic}. The automation of these tasks is often more closely related to signal processing than to artificial intelligence, but nonetheless it often involves a large component of machine intelligence, such as analyzing the internal structure of music \cite{paulus2010state}, recovering shared influences among performers \cite{liebman2012phylogenetic}, or identifying performers by nuances in their performance \cite{lagrange2012robust}. 		
		
A good example for a musical skill task, or music understanding task, is \emph{music segmentation}. Music segmentation strives to understand the structure of music by partitioning it into functionally separate and semantically meaningful segments. This partitioning can happen on multiple levels - a song could be partitions into an intro, verse, chorus, bridge, and outro for instance, and musical segments can be further broken down into independent musical phrases. The notion of recovering the rules of musical temporal structure is as old as musicology itself, and computational approaches to it date back to the work of Jackendoff and Lerdahl, who proposed a generative theory of tonal music in the early 1980s \cite{lerdahl1985generative}. In the modern computational research literature, early examples include the work of Batlle and Cano, who used Hidden Markov Models to identify boundaries in music sequences \cite{batlle2000automatic}, and Harford, who used self organizing maps for the same purpose \cite{harford2003automatic}. Similarly to Batlle and Cano, Sheh et al.\ also applied HMMs to segment chord sequences \cite{sheh2003chord}. Unlike Batlle and Cano, their approach is unsupervised - the most likely segmentation is extracted using the expectation-maximization (EM) method. Parry and Essa studied feature weighting for automatic segmentation, combining both local and global contour patterns to recover boundaries between musical phrases \cite{parry2004feature}. Liang et al.\ used Gaussian models to hierarchically segment musical sequences as a preprocessing step for classification \cite{liang2005hierarchical}. Pearce et al.\ compared statistical and rule-based models for melodic segmentation, achieving accuracy of nearly 87\% with a hybrid approach \cite{pearce2008comparison}. This work was interesting because it revealed (at the time) that data driven approaches alone underperformed compared to a method that combined both statistical boundary prediction and rule-based heuristics that incoporated preexisting knowledge of music theory.  Considering the harder problem of segmenting non-professional (and therefore messier and harder to process) recordings, Mueller et al.\ employed heuristic rules to segment raw recordings of folk tunes to individual notes in order to align them with MIDI versions \cite{muller2009robust}. To achieve this alignment, the audio was segmented in reference to the much neater MIDI input using a distance function that measures the distance between the chroma expected from the MIDI and those observed in the recording, thus accounting for potential deviations  in the non-professional performance. 

In strongly related work, Praetzlich and Mueller applied dynamic time warping to segment real opera recordings based on aligning them with a symbolic representation \cite{pratzlich2013freischutz}. In a different work, the same authors used techniques from the string matching literature to identify segments in recordings on a frame-level similarity basis \cite{pratzlich2014frame}. From a probabilistic perspective, Marlot studied a similar type of recordings made by amateur folk musicians, and trained a probabilistic model to segment them into phrases \cite{Marolt2009ProbabilisticSA}. In Marlot's approach, the signal is first partitioned into fragments that are classified into one of the following categories: speech, solo singing, choir singing, and instrumental music. Then, candidate segment boundaries are obtained
by observing how the energy of the signal and its content change. Lastly, Maximum aposteriori inference is applied to find the most likely set of boundaries (the training and evaluation were supervised and were done against a set of 30 hand-annotated folk music recordings). In more recent work, Rodriguez-Lopez et al.\ combined cue models with probabilistic approaches for melodic segmentation \cite{rodriguez2014multi}. Interestingly, in a paper from recent years, Lukashevich compared multiple possible metrics for song segmentation accuracy (a work also related to structure analysis, which we discuss in greater detail later in this subsection) \cite{lukashevichtowards}. In this work she exposed the fact that performance of different approaches can vary significantly when altering the accuracy metric. The somewhat subjective character of this task is also evident in the work of Pearce et al.



Along the same lines, much work has been invested in the tasks of \emph{chord extraction and harmonic modeling}, the practice of extracting the harmonic properties of a musical sequence, and reducing it to a more abstract representation of typical patterns. This task is of interest both from a general music understanding perspective and for practical applications such as music recommendation and preference modeling. The literature in this subfield has evolved in an interesting manner. Initial modern approaches, such as that of Paiement et al., were based on graphical models. Paiement et al.\ trained a graphical probabilistic model of chord progressions and showed it was able to capture meaningful harmonic information based on a small sample of recordings \cite{paiement2005probabilistic}. Burgoyne et al.\ compared multiple approaches of sequence modeling for automatic chord recognition, mainly comparing Dirichlet-based HMMs and conditional random fields (CRFs) over pitch class profiles \cite{burgoyne2005learning}. In something of a departure from the earlier problem setting, Mauch and Dixon used structural information about the music to better inform chord extraction, and utilized a discrete probabilistic mixture model for chord recognition, reaching average accuracy of ~65\% \cite{mauch2009using}. Cho and Bello introduced recurrence plots (essentially a derivative of the previously discussed self-similarity matrices) as a noise reduction method in order to smooth features and facilitate more accurate chord recognition, improving performance over a non-smoothed baseline. 

Unlike the probabilistic graphical models approach, Ogihara and Li trained N-gram chord models for the ultimate purpose of composer style classification (basically treating chords as words) \cite{ogihara2008n}. Combining the N-gram and probabilistic perspectives, Yoshii and Goto introduced a vocabulary free, infinity-gram model composite generative model for nonparametric chord progression analysis, which was able to recover complex chord progressions with high probability \cite{yoshii2010infinite}. Chen et al.\ expanded the standard HMM approach to chord recognition using duration-explicit HMM models \cite{chen2012chord}. Among their innovations is the utilization of a transformation matrix for chroma (learned via regression) that yields a richer spectral representation than that of the traditional chroma vector. On top of this learned representation a generalized, duration-aware HMM is used to predict the most likely chord sequence (using the Viterbi algorithm \cite{rabiner1989tutorial}). Papadopoulos and Tzanetakis chose to combine graphical models with a rule-based approach directly by utilizing a Markov logic networks to simultaneous model chord and key structure in musical pieces. More recently, deep neural networks have become increasingly prevalent for the purpose of chord recognition. Boulanger-Lewandowski et al.\ studied the application of recurrent neural networks (RNN), and specifically Restricted Boltzmann Machines (RBMs), for audio chord recognition \cite{boulanger2013audio}, and Humphrey and Bello applied convolutional neural networks (CNN) for the same purpose \cite{humphrey2012rethinking}. In a strongly  related paper, Zhou and Lerch trained a Deconvolutional neural networks (DNN) for feature construction, and combined SVM and HMM classifiers on a bottleneck layer of the DNN for final chord classification \cite{zhou2015chord}.



The problem of chord extraction and harmonic modeling is closely linked to that of \emph{note transcription and melody extraction}. Note transcription involves the translation of audio information into a sequential symbolic representation. Melody extraction is the related task of identifying a melodic sequence in a larger musical context and isolating it. Abdallah and Plumbley applied non-negative sparse coding \cite{hoyer2002non} on audio power spectra for polyphonic music transcription \cite{abdallah2004polyphonic}. Similarly, Ben Yakar et al.\ applied unsupervised bilevel sparse models for the same purpose \cite{yakar2013bilevel}. Madsen and Widmer introduced a formal computational model for melody recognition using a sliding window approach \cite{madsen2007towards}. In their work, they compared entropy measures with a compression-based approach to predict melody notes. Polliner and Ellis framed the melody transcription task as a classification problem, identifying notes in each frame based on the audio spectral properties \cite{poliner2006discriminative}. From a more statistical perspective, Duan and Temperley apply maximum likelihood sampling to reach note-level music transcription in polyphonic music \cite{duan2014note}. Alternatively, taking a Bayesian filtering approach, Jo and Yoo employed particle filters to track melodic lines in polyphonic audio recordings \cite{jo2010melody}. Kapanci and Pfeffer treated the melody extraction problem from an audio-to-score matching perspective, and trained a graphical model to align an audio recording to a score, recovering melodic lines in the process \cite{kapanci2005signal}. A different graphical approach to the problem was introduced by Raczynski et al., who trained a dynamic Bayes network (DBN) for multiple pitch transcription \cite{raczynski2010multiple}. In their study they were able to show this choice significantly improved performance compared to a reference model that assumed uniform and independently distributed notes.  Grindlay and Ellis propose a general probabilistic model suitable for transcribing single-channel audio recordings containing multiple polyphonic sources \cite{grindlay2010probabilistic}. As in other related problems, in the last few years multiple researchers have applied deep neural network architectures for this task. Boulanger-Lewandowski et al.\ applied RNNs to recover multiple temporal dependencies in polyphonic music for the purpose of transcription \cite{boulanger2012modeling}. Connecting the graphical model literature with the deep architectures thread, Nam et al.\ applied deep belief networks for unsupervised learning of features later used in piano transcription, showing an improvement over hand designed features \cite{nam2011classification}. In another recent work on piano transcription, Bock and Schedl applied bidirectional Long Short Term Memory RNNs (LSTMs), reporting improved performance compared to their respective baselines \cite{bock2012polyphonic}. Berg-Kirkpatrick et al.\ achieved the same goal of piano note transcription in a fully unsupervised manner, using a graphical model that reflects the process by which musical events trigger perceived acoustic signals \cite{berg2014unsupervised}. In another recent example, Sigtia et al.\ presented an RNN-based music sequence model \cite{sigtia2014rnn}. In the transcription process, prior information from the music sequence model is incorporated as a Dirichlet prior, leading to a hybrid architecture that yields improved transcription accuracy.







%

Chord analysis, melody extraction and music similarity are all strongly connected to \emph{cover song identification} - another field of music analysis where AI has been applied. Cover song identification is the challenging task of identifying an alternative version of a previous musical piece, even though it may differ substantially in timbre, tempo, structure, and even fundamental aspects relating to the harmony and melody of the song. The term ``cover'' is so wide that it ranges from acoustic renditions of a previous song, to Jimi Hendrix' famous (and radical) reinterpretation of Bob Dylan's ``All Along the Watchtower'', to Rage Against the Machine essentially rewriting Bob Dylan's ``Maggie's Farm''. Beyond its value for computational musicology and for enhancing music recommendation, cover song identification is of interest because of its potential for benchmarking other music similarity and retrieval algorithms. Ellis proposed an approach based on cross-correlation of chroma vector sequences, while accounting for various transpositions \cite{ellis2006identifying}. As a critical preprocessing step, chroma vectors were beat-aligned via beat tracking, a separate music information retrieval problem that we discuss further in this section. Serra et al.\ studied the application of Harmonic Pitch Class Profiles (HPCP) \cite{gomezHPCP, lee2006automatic} and local alignment via the Smith-Waterman algorithm, commonly used for local sequence alignment in computational biology \cite{smith1981comparison}, for this purpose \cite{serra2008chroma}. HPCP is an enhancement of chroma vectors which utilizes the typical overtone properties of most instruments and the human voice to obtain a less noisy representation of the pitch class profile of a musical segment. Serra at el. later proposed extracting recurrence measures from the cross recurrence plot, a cross-similarity matrix of beat-aligned HPCP sequences, for more accurate cover song identification. Since complicated pairwise comparisons for the purpose of en masse cover song identification in large scale datasets is prohibitively computationally expensive, Bertin-Mahieux and Ellis proposed a significant speed-up to previous approaches by extracting the magnitude of the two-dimensional Fourier transform of beat-aligned chroma patches (chroma patches are windowed subsequences of chroma vectors) and then computing the pairwise euclidean distance of these representations (PCA was also applied for dimensionality reduction) \cite{bertin2012large}. Humphrey et al.\ further improved on this result by introducing various data-driven modifications to the original framework. These modifications included the application of non-linear scaling and normalization on the raw input, learning a sparse representation, or a dictionary (essentially a set of approximate basis functions that can be used to describe spectral patterns efficiently) in order to further reduce the complexity of the input data \cite{humphrey2013data}. More recently, Tralie and Bendiche cast the cover song identification problem as matching similar yet potentially offset, scaled and rotated patterns in high-dimensional spaces, treating MFCC representations as point-cloud embeddings representing songs \cite{tralie2015cover}.


Another important aspect of computational music analysis where machine intelligence has been applied is that of \emph{onset detection}. Onset detection refers to the issue of identifying the beginning of notes in audio representations, and it has been widely studied given its fundamental application to music information analysis. You and Dannenberg proposed a semi-supervised scheme for onset detection in massively polyphonic music, in which more straightforward signal processing techniques such as thresholding, are likely to fail due to the difficulty in disambiguating multiple adjacent notes with overlapping spectral profiles \cite{you2007polyphonic}. To avoid the necessity of hand labeling the countless onsets, audio-to-score alignment is used to estimate note onsets automatically. Because score alignment is done via chroma vectors, which only provide crude temporal estimates (on the order of 50 to 250ms), a trained support vector machine classifier is used to refine these results. Later, Benetos et al.\ showed that using the auditory spectrum representation can significantly improve onset detection \cite{benetos2009pitched}. Inspired by both computational and psycho-acoustical studied of the human auditory cortex, the auditory spectrum model consists of two stages, a spectral estimation model (designed to mimic the cochlea in the
auditory system), and a spectral analysis model. Extracting the group delay (the derivative of phase over frequency) \cite{holzapfel2008beat} and spectral flux (the detection of sudden positive energy changes in the signal) \cite{bello2005tutorial}, the authors were able to reach dramatic improvements in performance compared to more straightforward Fourier-based onset detection \cite{benetos2009pitched}. More recently, Schluter and Bock were able to significantly improve on previous results by training a convolutional neural network for the purpose of beat onset detection \cite{schluter2014improved}.

The notion of onset detection naturally leads to another core property of music that has been studied computationally - \emph{beat perception}. The beat of a musical piece is its basic unit of time. More concretely, by ``beat perception'' we refer to the detection of sequences of temporal emphases that induce the perceived rhythm of a musical piece. We have touched on the issue of beat detection explicitly when we discussed cover song identification (when discussing the works of Ellis et al.~\cite{ellis2006identifying} and Serra et al.~\cite{serra2008chroma}), but in truth the issue of beat tracking is present in almost any task that involves the comparative analysis of audio sequences (in symbolic representations the issue of beat tracking is significantly less challenging for obvious reasons). Raphael introduced a generative model that captures the simultaneous nature of rhythm, tempo and observable beat processes and utilized it for automatic beat transcription. Given a sequence of onset times, a sequence of measure positions, and a Gaussian tempo process, a graphical model is used to describe the process with which these sequences are connected. Using maximum aposteriori inference, the sequence of beats is produced \cite{raphael2001automated}.  Alonso et al.\ defined the notion of spectral energy flux (which we mentioned previously in the context of onset detection) to approximate the derivative of the energy per frequency over time, and use it for efficient beat detection \cite{alonso2004tempo}. Paulus and Klapuri combine temporal and spectral features in an HMM-based system for drum transcription \cite{paulus2007combining}. Temporal patterns are modeled as a Gaussian Mixture Model, and are combined with a hidden Markov Model that considers the different drum combinations, and the drum sequence is inferred via maximum likelihood. Gillet and Richard also tackled drum transcription specifically, but took a different approach, training a supervised N-gram model for interval sequences \cite{gillet2007supervised}. In their method, after extracting initial predictions based on the N-gram model, a pruning stage takes place in an unsupervised fashion, by reducing the approximate Kolmogorov complexity of the drum sequence. Le Coz et al.\ proposed a different approach altogether to beat extraction, which does not rely on onset detection, but rather on segmentation \cite{le2010segmentation}. In their paper, they segment each note into quasi-stationary segments reflecting (approximately) the attack, decay, sustain and release of the notes via forward-backward divergence \cite{andre1988new}, and reconstruct the beat sequence directly from the resulting impulse train via Fourier analysis. 

Beat extraction is closely related to \emph{audio-to-score alignment} and score following - the task of matching audio to a score in an online fashion (we have already touched on this subject in the context of melody extraction and onset detection). Dixon proposed an application of the Dynamic Time Warping algorithm for this purpose \cite{dixon2005line}. Dynamic Time Warping is a well known dynamic programming algorithm for finding patterns in time series data by aligning two time-dependent sequences \cite{berndt1994using}, and its application in the context of aligning scores to audio data is self-evident (it was also used context such as cover song identification, which we have discussed previously). Pardo and Birmingham tackled the score following from a probabilistic perspective \cite{pardo2005modeling}. In their paper, they treating the score as a hidden Markov model, with the audio as the observation sequence, reducing the score following to the problem of finding the most likely state at a given point, which can be done via Viterbi-style dynamic programming. In a recent paper, Coca and Zhao employed network analysis tools to recover rhythmic motifs (represented as highly connected graph sub-components) from MIDI representations of popular songs \cite{coca2016musical}.






Melody, harmony and rhythm modeling, and score alignment, all naturally lead to the task of overall \emph{musical structure analysis}. This problem has been studied as well, from multiple directions. Kameoka et al.\ employed expectation-maximization to recover the harmonic-temporal overall structure of a given piece. Abdallah et al.\ propose a Bayesian approach to clustering segments based on harmony, rhythm, pitch and timbre. Peeters applies spectral analysis to the signal envelope to recover the beat properties of recorded music \cite{peeters2007sequence}. Peeters' approach was to utilize MFCC and pitch class profile features, construct higher order similarity matrices, and infer the structure via maximum likelihood inference. Mueller and Ewert jointly analyze the structure of multiple aligned versions of the same piece to improve both efficiency and accuracy \cite{muller2008joint}. This type of analysis is done by finding paths in the pairwise similarity matrix of chroma vector sequences and using them to partially synchronize subsequences in both pieces. Bergeron and Conklin designed a framework for encoding and recovering polyphonic patterns in order to analyze the temporal relations in polyphonic music \cite{bergeron2008structured}. To achieve this sort of encoding, they proposed a polyphonic pattern language inspired by algebraic representations of music, which can be seen as a formal logic derivation system for harmonic progressions. From a more utilitarian perspective, as an example for structure analysis as a preprocessing step for other purposes, Mauch et al.\ used patterns recovered from music structure to enhance chord transcription. Harmonic progressions in Western music tend to obey contextual and structural properties (consider, for instance, the cadenza, a typical harmonic progression signifying the end of a musical phrase). Specifically, in their work, Mauch et al.\ leverage repetitions in sequences to improve chord extraction by segmenting the raw sequence and identifying those repetitions. From a different perspective, Kaiser and Sikora used nonnegative matrix factorization to recover structure in audio signals \cite{kaiser2010music}. The nonnegative matrix factorization is applied on the timbre self-similarity matrix, and regions of acoustically similar frames in the sequence are segmented. Another unsupervised approach for overall structure analysis is described in more recent work by McFee and Ellis, who employed spectral clustering to analyze song structure. They construct a binary version of the self-similarity matrix which is subsequently interpreted as a unweighted, undirected graph, whose vertices correspond to samples. Then, spectral clustering (through Laplacian decomposition) is applied, with the eigenvalues corresponding to a hierarchy of self-similar segments. In a somewhat related recent paper, Masden et al learned a pairwise distance metric between segments to predict temporally-dependent emotional content in music \cite{madsen2014modeling}.

A research topic that is related to structure analysis, beat perception, melody, and chord extraction is that of \emph{motive identification} - the extraction of key thematic subject matter from a musical piece. To mention a few papers from the past 15 years, Juhasz studied the application of self-organizing maps and dynamic time warping for the purpose of identifying motives in a corpus of 22 folk songs \cite{juhasz2009motive}. Dynamic time warping is used to search for repeated subsequences in melodies (in a way conceptually related to how self-similarity matrices work), and then these sequences are fed to a self organizing map, extracting the most prominent abstracted representations of the core motifs and their correspondence relationships. Lartillot framed the motive extraction problem as combinatorially identifying repeated subsequences in a computationally efficient manner \cite{lartillot2005efficient}. The subsequences is multidimensional, as it comprises both melodic and rhythmic properties. Lartillot later revisited and refined this approach, and tested in on the Johannes Kepler University Patterns Development Database \cite{collins2013discovery}, and was able to show it recovers meaningful motivic patterns.

Lastly, it is worth mentioning another example for the application of AI towards musicological problems - \emph{performance analysis}. The rise in corpora of recorded music has both facilitated and necessitated the application of algorithmic approaches to comparatively analyze multiple recordings of the same pieces. Several good examples for such computational method include the work of Madsen and Widmer, who applied string matching techniques to compare pianist styles \cite{madsen2006exploring}. In a related work, Sapp used rank similarity matrices for the purpose of grouping different performances by similarity \cite{sapp2007comparative}. Molina-Solana et al.\ introduced a computational experssiveness model in order to improve individual violinist identification \cite{molina2008using}. In past work, Liebman et al.\ applied an approach inspired by computational bioinformatics to analyze the evolution and interrelations between different performance schools by constructing an evolutionary tree of influence between performances \cite{liebman2012phylogenetic}. Other related works include that of Okomura et al., who employed stochastic modeling of performances to produce an ``expressive representation'' \cite{okumura2011stochastic}. More recently, van Herwaarden et al.\ trained multiple Restricted Boltzmann Machines (RBMs) to predict expressive dynamics in piano recordings \cite{van2014predicting}.





		
		\subsection{Generation Tasks}
		\label{chap8:gen}
		
Thus far we have considered tasks where intelligent software was required to perform tasks with existing pieces of music as input. However, there is also a wide array of work on employing artificial agents for the purpose of creating music. The autonomous aspect of algorithmic composition has been routinely explored in various artistic contexts \cite{nierhaus2009algorithmic}. However, while considered by some as the ``holy grail'' in computer music and the application of AI to music, less scientific attention has been placed on AI for musical content generation compared to other music AI problems.\footnote{By ``scientific'' we primarily mean principled, measurable and reproducible research in appropriate publication venues.} This gap owes at least in part to the fact that evaluating the quality of computer generated content is very difficult, for reasons discussed in Section \ref{chap8:eval}


In many ways, the task of \emph{playlist generation}, or recommending music in a sequential and context dependent manner, can be perceived as lying at the intersection of recommendation and generation. In the past 15 years, multiple works have studied machine learning approaches to created meaningful song sequences. Maillet et al.~\cite{maillet2009steerable} treated the playlist prediction problem as a supervised binary classification task, with pairs of songs in sequence as positive examples and random pairs as negative ones. Mcfee and Lanckriet~\cite{mcfee2011natural} examined playlists as a natural language model induced over songs, and trained a bigram model for transitions. Chen et al.~\cite{chen2012playlist} took a similar Markov approach, treating playlists as Markov chains in some latent space, and learned a metric representation for each song without reliance on audio data. Zheleva et al.~\cite{zheleva2010statistical} adapted a Latent Dirichlet Allocation model to capture music taste from listening activities across users and songs. Liebman et al.~\cite{DJMC} borrow from the reinforcement learning literature and learn a model both for song and transition preferences, then employing a monte-carlo search approach to generate song sequences. Wang et al.~\cite{wang2013exploration} consider the problem of song recommendations as a bandit problem, attempting to efficiently balance exploration and exploitation to identify novel songs in the playlist generation process, and very similar work has been done by Xing et al.~\cite{xing2014enhancing} towards this purpose as well. Novelty and diversity in themselves have also been a studied objective of playlists. Logan and Salomon \cite{logan2001music, logan2002content} considered novelty in song trajectories via a measure which captures how similar songs are from one another in a spectral sense. Lehtiniemi \cite{lehtiniemi2008evaluating} used context-aware cues to better tailor a mobile music streaming service to user needs, and showed that using such cues increases the novelty experienced by users. More recently, Taramigkou et al.~\cite{taramigkou2013escape} used a combination of Latent Dirichlet Allocation with graph search to produce more diversified playlists that are not pigeonholed to overly specific tastes, leading to user fatigue and disinterest. 
		

Another task of a generative nature is that of \emph{expressive performance}. It is naturally closely related to music performance analysis, but rather than perceiving how humans perform music expressively, the emphasis in this task is on imparting computational entities with the ability to generate music that would seem expressive to a human ear. Early modern approaches to this problem include the work of de Mantaras et al., who applied case-based reasoning for the purpose of algorithmic music performance \cite{de2002ai}, and that of Ramirez and Hazan, who used a combination of k-means clustering and classification trees to generate expressive performances of Jazz standards \cite{ramirez2006tool}. Ramirez et al.\ later proposed a sequential covering evolutionary algorithm to train a model of performance expressiveness based on Jazz recordings \cite{ramirez2007inducing}. Diakopoulos et al.\ proposed an approach for classifying and modeling expressiveness in electronic music, which could also be harnessed for generating automatic performances \cite{diakopoulos200921st}.

The challenge of expressive performance has been of particular interest in robotic platforms. Murata et al.\ studied the creation of a robotic singer which was able to follow real-time accompaniment \cite{murata2008robot}. In a somewhat related paper, Xia et al.\ presented a robotic dancer which tracked music in real time and was trained to match the expressiveness of the music with matching dance movement \cite{xia2012autonomous}. Another example is the work of Hoffman and Weinberg, who presented Shimon, a robotic marimba player, and borrowed ideas from the world of animation to make Shimon expressive not just musically, but also visually \cite{shimon}.

%

Shimon was geared towards \emph{live improvisation}, and indeed improvisation is yet another music generation goal for artificial systems. Eck and Schmidhuber used long short-term memory recurrent neural networks to train a generative model of Jazz improvisation \cite{eck2002finding}. In a different contemporary work, Thom employed a learned probabilistic model for interactive solo improvisation with an artificial agent \cite{thom2001machine,thom2000unsupervised}. Assayag and Dubnov trained Markov models for music sequences, then employ a type of string matching structures called factor oracles to facilitate algorithmic improvisation \cite{assayag2004using}. 

Lastly, there has been some attention from an AI perspective on automatic music generation, though the study of this problem has been relatively limited, particularly due to the difficulty of evaluation (see Section \ref{chap8:eval}). In a technical report, Quick borrowed  ideas from Shenkerian analysis and chord spaces to create an algorithmic composition framework \cite{quick2010generating}. Kosta et al.\ proposed an unsupervised multi-stage framework for chord sequence generation based on observed examples \cite{kosta2012unsupervised}. From a very different perspective, Blackwell has applied multi-swarms to create an improvisational musical system \cite{blackwell2003swarm}. Very recently, Colombo et al.\ proposed deep RNN architectures for the purpose of melody composition \cite{colombo2016algorithmic}. Most recently, Dieleman et al.\ compared different deep architectures for generating music in raw audio format at scale \cite{dieleman2018challenge}, and Huang et al.\ were able to apply deep sequential generative models with self-attention to generate structured compositions that achieve state of the art performance in synthesizing keyboard compositions \cite{huang_transformer}. Similarly, quite recently, Payne proposed MuseNet, a deep neural network model that can generate several minutes long compositions for ensembles of up to ten different instruments, reasoning about musical styles in the process \cite{musenet}. For an interesting overview of AI methods particularly in the use of algorithmic composition, see \cite{fernandez2013ai}.



%
%
%
%

%

	\section{Overview of Common Representations}
\label{chap8:repr}
Thus far, we have focused on breaking down the wide range of musical tasks from a purpose-oriented perspective. However, an equally important perspective involves the types of input used for these tasks. As noted by Dannenberg \cite{dannenberg1993music}, representation of the music itself can be viewed as a continuum ``ranging from the highly symbolic and abstract level denoted by printed music to the non-symbolic and concrete level of an audio signal''. Additionally, one may consider all the additional related information, such as lyrics, tags, artist's identity, etc.\ as part of the representation. As briefly mentioned in Section \ref{chap8:tax}, we consider three main types of information categories for music:

\begin{itemize} 
\item Symbolic representations - logical data structures representing musical events in time, which may vary in level of abstraction. Examples for different levels of abstraction include but are not limited to the extent of encoded detail regarding pitch, registration, timbre, and performance instructions (accents, slurs, etc).

\item Audio representations - this sort of representation captures the other end of the continuum mentioned above, capturing the audio signal itself. Despite its seeming simplicity, here too there is a level of nuance, encompassing the fidelity of the recording (levels of compression, amplitude discretization and so forth), or the level of finesse in representations which perform signal processing on the original audio (such as the ubiquitous chroma and MFCC audio representations we have already mentioned in Section \ref{chap8:tasks} and discuss in further detail later in this section).

\item Meta-musical information - all the complementary information that can still be legitimately considered part of the musical piece (genre classification, composer identity, structural annotations, social media tags, lyrics etc). 
\end{itemize}

\begin{figure}[!htb]
\centering
\includegraphics[width=.8\linewidth]{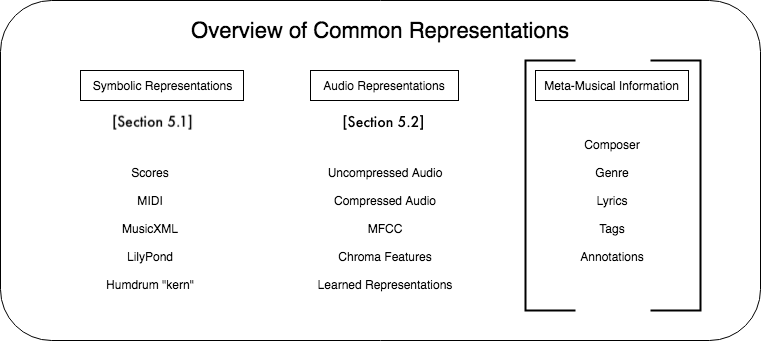}
\caption{Visual high-level overview of music representations used in music AI research. For reasons described in the text, we only consider the first two categories in this article.} \label{chap8:fig_repr}
\end{figure}

Of these three broad categories, only the first two are within the scope of this survey, since we explicitly focus on aspects of music analysis relating to the music itself, rather than applying machine learning directly and/or exclusively on the complementary information such as lyrics, social media context, or general artist profiles. A visual summary of the contents of this section is presented in Figure \ref{chap8:fig_repr}.

 We now expand on the first two types of input.

		\subsection{Symbolic Representations for Music}
		
One of the earliest and most common approaches to representing music inputs is via symbolic formats. In essence, a symbolic representation of music is the conceptual abstraction, or the blueprint, of that music. Musical scores using Western notation, for instance, serve exactly as such blueprints. In its most basic form, it includes information on pitches, their length, and when they are played. Additional relevant information includes when each note is released, the amplitude of each note, and the attack (simply put, how rapidly the initial rise in amplitude is and how amplitude decays over time). Classical scores also include a wide range of additional data regarding performance, such as performance instructions, sound effects, slurs, accents, and so forth, all of which can often be represented in symbolic formats as well. Additional information such as timbre can be represented, typically by using a preexisting bank of instrument representations. 

While this representation isn't as rich as an audio recording, for certain genres, such as classical music or musical theater, which already rely on scores, it is an incredibly informative and useful resource, that eliminates multiple levels of difficulty in dealing with complex auditory data, enabling an artificial agent to know at each moment the core information about pitch, dynamics, rhythm and instrumentation.

One of the most common ``blueprint'' formats is the MIDI protocol. Since its initial introduction in the early '80s, the MIDI (Musical Instrument Digital Interface) format has served as part of the control protocol and interface between computers and musical instruments \cite{loy1985musicians}. The MIDI format specifies individual notes as ``events'' represented as tuples of numbers describing varied properties of the note including pitch, velocity (amplitude), vibrato and panning. These note events are sequenced to construct a complete piece, containing up to 16 separate channels of information. These channels typically represent instruments, since each channel can be associated with a separate sound profile, but sometimes the same instrument can be partitioned into multiple channels. Due to its long history and ubiquity, much of the literature utilized this file format as input source (See \cite{rizo2006pattern,yeshurun2006midi,yang2017midinet,grosche2010makes,rauber2002using,madsen2007towards,hillewaere2010string,tsai2005query,anan2012polyphonic,mardirossian2006music} for a very partial list of examples).

A different approach to symbolic representation aims to digitally represent musical scores, similarly to how traditional music engraving generates scores for mass printing. In the past two decades, several such formats have emerged, including LilyPond \cite{nienhuys2003lilypond}, Humdrum ``kern'' \cite{huron2002music,sapp2005online} and MusicXML \cite{good2001musicxml}, among others. While this list is not comprehensive, in terms of symbolic music analysis these formats are largely equivalent and can be converted from one to another with some loss of nuance, but preserving most key features. Examples of research utilizing data in these formats is plentiful and varied (see \cite{sapp2005online,sinclair2006lilypond,cuthbert2011feature,antila2014vis}, for, once again, a very partial list of examples). 

The advantage of using such music engraving representations, particularly from a musicology perspective, is that they are designed to capture the subtleties of Western notation, including concepts such as notes, rests, key and time signatures, articulation, ornaments, codas and repetitions, etc. This richness of representation is in contrast to the MIDI format, which is conceptually closer to raw audio in terms of abstractions and is designed to describe specific pitched events in time, and is thus less suited to capture the full complexity of more sophisticated music scoring. On the flipside, that is also the relative strength of MIDI compared to these other formats - it is much simpler to parse and process. Furthermore, from a practical standpoint, MIDI largely predates these other formats and is designed as an interface protocol rather than a music engraving language, and is thus far more commonly supported by electronic musical instruments, devices, and software.


		\subsection{Audio Representations and Derived Features}

A more intuitive way to represent music is through digital sampling of the raw audio, as is done on audio CDs and using the canonical wave and aiff file formats. In its crudest form, digitizing music audio simply captures amplitude over time in either 
a single (mono) or dual (stereo) output channel. The quality of recording is dependent on 
two main aspects:
\begin{itemize}
\item The number of bits used to represent amplitudes, which 
determines quantization noise.
\item The sampling frequency, which determines the range of frequencies captured in the digitization process. The standard 
sampling frequency of 44100Hz ensures that no human audible frequencies 
are lost.
\end{itemize}

To these considerations one may also add the possibility of using compression, typically at some cost to frequency resolution \cite{pye2000content}. Historically, working directly on raw audio has proven impractical. First, it has traditionally been prohibitively expensive in terms of data storage and processing cost. Second, and more importantly, it has been impractical in terms of the ability of AI software to extract meaningful information from such a low level representation. For reference, this pattern is somewhat analogous to the historical difficulty in using raw pixel data in visual processing. 

For this reason, similar to how visual processing resorted to more expressive, condensed representations such as SIFT \cite{lowe1999object} and HOG \cite{dalal2005histograms} features, different features constructed from raw audio have been commonly used. The common are the Mel-frequency cepstral coefficients (MFCC) \cite{logan2000mel}, a derivative of the Fourier transform which captures the short-term power spectrum of a sound. The MFCC is typically constructed using successive temporal windows, thus representing auditory information over time. These coefficients were first used in speech recognition \cite{hasan2004speaker}, and over the past two decades were shown to be extremely useful in music analysis, serving as a condensed but expressive representation of spectrum over time (see \cite{proutskova2009you,schuller2010vocalist,tomasik2009using,han2014hierarchical,marolt2009probabilistic} for a few examples). 

To reiterate, the symbolic and the auditory aspects of music representation aren't separate categories but rather the two ends of a continuum. A good example for a commonly used representation that lies somewhere in between these two ends is that of chroma features \cite{ellis2007identifyingcover}. As we've briefly mentioned in Section \ref{chap8:tasks}, chroma features record the intensity associated with each of the 12 semitones in an octave, thus, when windowed, capture both melodic and harmonic information over time. Since this representation is typically extracted via analyzing the spectrum of the music, and since it strives to achieve a succinct representation of the notes physically heard throughout a recording, it has something of the auditory representation. At the same time, it also reduces raw audio to a series of pitch information over time, thus also retaining something of the symbolic. 

There is an inherent trade-off in choosing a music representation. Audio information is ubiquitous and more immediately useful for large-scale common applications. At the same time, raw recordings are harder to analyze, store and query. Symbolic representations are elegantly concise ways of storing and relaying a great deal of the audio information Western music traditionally cares about (which is in part why reading sheet music is still considered a fundamental skill for musicians), and such representations can be used efficiently for many analysis and retrieval tasks, but they are generally less common, less valuable for mass use and inherently partial in the sense that ultimately crucial auditory information is nonetheless lost. In practice, the choice of representation in the literature is more often than not dictated by availability, ease of use and the nature of the studied task.

In the past few years, as part of the rising popularity and success of deep learning \cite{lecun2015deep}, multiple papers have explored the prospects of using deep artificial neural networks to autonomously learn representations - i.e., learn meaningful features - from raw audio. Lee at al. \cite{lee2009unsupervised} have shown that generic audio classification features learned using convolutional deep belief networks were also useful in 5-way genre classification. Hamel and Eck also explored deep belief nets for both genre classification and automatic tagging, and have shown their learned features to outperform the standard MFCC features \cite{hamel2010learning}. Henaff et al.\ used sparse coding to learn audio features and showed this approach to be competitive with the state of the art in genre classification on a commonly used dataset \cite{henaff2011unsupervised}. Humphrey et al.\ surveyed various aspects of deep feature learning, and analyzed how the proposed architectures can be seen as powerful extensions for previously existing approaches \cite{humphrey2012moving}. While these new approaches are certainly promising, such architectures have not fully supplanted the previously designed representations discussed in this section, and are not a replacement for existing music interface protocols such as MIDI and music-engraving languages such as LilyPond.


	\section{Overview of Technique}
	\label{chap8:technique}
\begin{figure}[!htb]
\centering
\includegraphics[width=.8\linewidth]{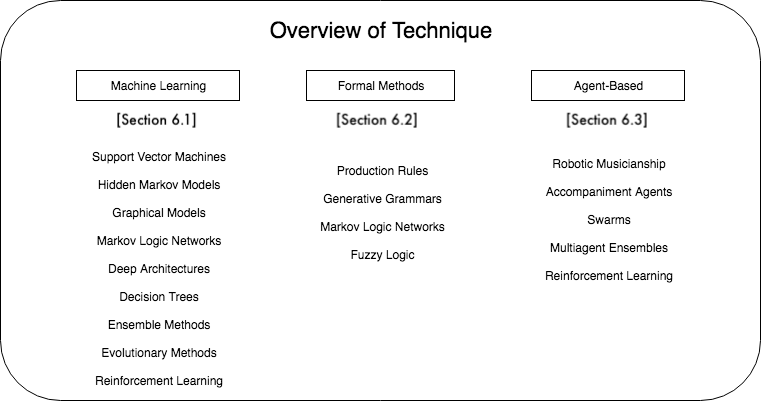}
\caption{Visual high-level overview of algorithmic techniques used in music AI research.} \label{chap8:fig_tech}
\end{figure}

A wide variety of machine learning and artificial intelligence paradigms and techniques have been applied in the context of music domains. From a machine learning and artificial intelligence research perspective, it is of interest then to examine this range of techniques and the specific musical domains where they were applied. Due to the extensive nature of the related literature and the wide range of musical tasks where the following methods have been used, this list cannot be entirely comprehensive. To the best of our knowledge, however, it is representative of the full array of methods employed. A visual summary of the contents of this section is presented in Figure \ref{chap8:fig_tech}.

			\subsection{Machine Learning Approaches}

Considering the long list of music informatics tasks described in section \ref{chap8:tasks}, it is clear that many of them can be viewed as machine learning problems. Indeed, a broad spectrum of machine learning techniques have been used to tackle them.

Perhaps one of the oldest documented machine learning approaches for musical tasks is \emph{support vector machines} (SVM) and kernel methods. As mentioned in Section \ref{chap8:classification}, in an early example of computational approaches to music in general, Marques and Moreno utilized SVM for instrument classification \cite{marques1999study}. Xu et al.\  used a multi-layer SVM approach for genre classification \cite{xu2003musical}. Their approach was to use the different features representing the spectrum of the audio and hierarchically partition the input first to Pop/Classic or Rock/Jazz, and then within each category (all in all training three SVM models). A similar task was also pursued by Mandel and Ellis, who studied the application of SVM on song-level features for music classification \cite{mandel2008multiple}. Meng and Shawe-Taylor studied other types of feature models, namely multivariate Gaussian models and multivariate autoregressive models, for short time window feature representation, with the ultimate goal of improved classification results over 11 genre categories \cite{shawe2005investigation}. Han et al.\ used the strongly related technique of support vector regression for emotion classification in music \cite{han2009smers}. Their proposed SMERS system extracts features from the raw audio, maps given audio from its feature representation to Thayer’s two-dimensional emotion
model (this emotion representation is based on, and trains a support vector regressor for future prediction.  Helen and Vitranen used support vector machines to classify audio components as drums vs. pitched instruments \cite{helen2005separation}. Ness et al.\ applied a stacked SVM approach for automatic music tagging, using the key insight that the probabilistic output of one SVM can be used as input for a second layer SVM in order to  exploit possible correlations between tags \cite{ness2009improving}. Maddage et al.\ trained an SVM classifier to distinguish purely instrumental music sections from ones mixing instruments and vocals, for the purpose of song structure analysis \cite{maddage2003svm}. Gruhne et al.\ used SVM classifiers for phoneme identification in sung lyrics in order to synchronize audio with text \cite{gruhne2007detecting}. While useful, the overall popularity of SVM approaches for music informatics seems to have somewhat faded in the past few years, perhaps reflecting its diminishing popularity in the machine learning community in general.


Another well-established and frequently used machine learning approach for musical tasks is that of \emph{probabilistic methods}. Standard examples include Hidden Markov Models (HMM), which are of obvious use given the sequential and partially observable nature of music. In early examples, Battle and Cano used competitive HMMs (or Co-HMMs), a variation on the standard HMM paradigm, for automatic music segmentation \cite{batlle2000automatic}. In their study, Co-HMMs were better suited for music partitioning since they required far less apriori domain-knowledge to perform well. Durey et al.\ used HMMs for the purpose of spotting melodies in music \cite{durey2001melody}, extracting notes from raw audio and treating them as observations in a graphical music language model. Eichner et al.\ were able to use HMMs for instrument classification. In their paper, they manually collected fragments of solo recordings of four instruments: classical guitar, violin, trumpet and
clarinet, and trained separate HMMs for each instrument, leveraging the fact that different instruments induce different note transition mechanics \cite{eichner2006instrument}. Sheh and Ellis used HMMs for the more complicated task of chord recognition and segmentation \cite{sheh2003chord}, while Noland and Sandler trained an HMM for key estimation \cite{noland2006key}. Extending these directions, Burgoyne and Saul applied a hidden Markov model to train Dirichlet distributions for major and minor keys on normalized pitch
class profile vectors, for the eventual purpose of tracking chords and keys over time \cite{burgoyne2005learning}. 

Chen et al.\ used a duration-explicit HMM (or DHMM) for better chord recognition \cite{chen2012chord}. DHMMs work in different time resolutions to estimate
the chord sequence by simultaneously estimating chord labels
and positions. In their paper, Chen et al.\ were able to show that explicitly modeling the duration of chords improved recognition accuracy. Considering a different approach, Papadopoulos and Tzanetakis applied Markov Logic Networks (MLNs) for modeling chord and key structure, connecting the probabilistic approach with logic-based reasoning \cite{papadopoulos2012modeling}. In practice, their approach is to take Markov networks that encode the transitional chord dynamics of particular scales and combine them with a first-order knowledge base that encodes rules such as ``A major chord implies a happy mood''. Leveraging the generative capabilities of HMMs, Morris et al.\ proposed a system that uses a Hidden Markov Model to generate chords to accompany a vocal melody \cite{morris2008exposing}. More recently, Nakamura et al.\ studied the application of autoregressive Hidden Semi-Markov Models for score following \cite{nakamura2015autoregressive}, as well as for recovering piano fingering \cite{nakamura2014merged}. In the context of ethnomusicology, Jancovic et al.\ applied HMMs for automatic transcription of traditional Irish flute music \cite{jancovic2015automatic}.

\emph{Graphical models} in general have been used in various ways in music domains. Raphael designed a graphical model for recognizing sung melodies \cite{raphael2005graphical} and for aligning polyphonic audio with musical scores \cite{raphael2004hybrid}. Kapanci and Pfeffer explored the related notion of graphical models for signal-to-score music transcription, modeling different aspects of the music such as rhythm and pitch as first-order Gaussian processes \cite{kapanci2005signal}. Pickens and Iliopoulos proposed a Markov Random Fields (MRFs) for general music information retrieval tasks \cite{pickens2005markov}, citing the power of MRFs in handling non-independent features as their key strength and inherently suitable for music tasks, in which various aspects of features - pitch, timbre, tempo etc) are deeply interdependent. Hoffman et al.\ used a hierarchical Dirichlet process to estimate music similarity \cite{hoffman2008content}. Hu and Saul proposed an approach a key profiling modeling technique that utilizes a latent Dirchilet allocation (LDA)  topic model \cite{hu2009probabilistic}. The core insight in their paper was that by looking for commonly cooccurring notes in songs, it is possible to learn distributions over pitches for each musical key individually. Yoshii and Goto proposed a novel model for spectral representation called infinite latent harmonic allocation models (iLHA) \cite{yoshii2010infinite}. Their model represents a Bayesian Nonparametric approach in which each spectral basis is parameterized by means of a Gaussian mixture model (GMM), with both the number of bases and the number of partials being potentially infinite (in practice the least informative elements are zeroed out quickly and a finite approximation remains). In their paper they show this model is useful for multipitch analysis. More recently, Berk-Kirkpatrick et al.\ proposed a graphical model for unsupervised transcription of piano music, designing a complicated probabilistic activation model for individual keystrokes and inferring the most plausible sequence of key activations to produce a given spectogram \cite{berg2014unsupervised}. Schmidt and Kim proposed a conditional random field (CRF) approach for tracking the emotional content of musical pieces over time \cite{schmidt2011modeling}. Later, the same authors would study the application of deep belief networks to learn better music representations, to be used later on in supervised learning tasks \cite{schmidt2013learning}. Another very current example of the application of deep generative models for musical task is the work of Manzelli et al., who applied a Long Short Term Memory network (commonly referred to as LSTMs) to learn the
melodic structure of different styles of music, and then use the unique symbolic generations from this model as a conditioning input for an audio generation model \cite{manzelli2018conditioning}. In a different recent work, Korzeniowski and Widmer proposed an RNN-based probabilistic model that allows for the integration
of chord-level language models with frame-level acoustic models, by connecting the two using chord duration models \cite{korzeniowski2018improved}.

As illustrated by these last few examples, the concept of deep belief networks and deep generative models in general is a natural bridge between graphical models and artificial neural network architectures, which indeed constitute the next learning paradigm we will discuss.


\emph{Artificial Neural Networks} (ANN) are among the oldest paradigms of machine learning. As such, they are also among the oldest to have been used by computational researchers studying musical tasks. To mention a several early modern examples, as early as 1997, Dannenberg et al.\ used ANNs, among other techniques, for musical style recognition \cite{dannenberg1997machine}. Kiernan proposed ANNs for score-based style recognition \cite{kiernan2000score}, and Rauber et al.\ applied a self-organizing map (SOM) on psycho-acoustic features to learn a visualization of music datasets \cite{rauber2002using}. For some additional details on the prehistory of this approach, it is worth reviewing Griffith and Todd's 1999 short survey on using ANNs for music tasks \cite{griffith1999musical}.

In recent years, after an extended lapse in popularity, there has been a resurgence for ANNs via \emph{deep architectures} (commonly dubbed ``deep learning''). Naturally, these learning architectures have also been firmly embraced by researchers at the intersection of AI and music. Boulanger-Lewandowski et al.\ studied audio chord recognition using Recurrent Neural Networks (RNNs) \cite{boulanger2013audio}. Herwaarden et al.\ applied Restricted Boltzmann Machines (RBMs) for predicting expressive dynamics in piano performances \cite{van2014predicting}. Bock and Schedl applied RNNs for automatic piano transcription \cite{bock2016joint} and for joint beat and downbeat tracking \cite{bock2016joint}. In the latter work,  an RNN operating directly on magnitude spectrograms is used to model the metrical structure of the audio
signals at multiple levels and provides an output feature for a Dynamic Bayes Network which models the bars, thus making this work another example for the fusion of deep architectures and graphical models. Krebs et al.\ also utilized RNNs for the purpose of downbeat tracking \cite{krebs2016downbeat}, using a very similar RNN + Dynamic Bayes Network learning framework, but in that work they used beat-synchronous audio features rather than the spectogram information. Humphrey et al.\ applied Convolutional Neural Networks (CNNs) for automatic chord recognition \cite{humphrey2012rethinking}. Humphrey has also been able to show the utility of deep architecture to learn better music representations \cite{humphrey2012moving}. CNNs were also recently used by Choi et al.\ for automatic tagging \cite{choi2016automatic}. In that paper, they use the raw mel-spectorgram as two-dimensional input, and compare the performance of different network architectures, and study their prediction accuracy over the MagnaTagATune dataset. Vogl et al.\ applied RNNs for automatic drum transcription, training their model to identify the onsets of percussive instruments based on general properties of their sound \cite{vogl2016recurrent}. Liu and Randall applied bidirectional Long Short Term Memory networks (LSTMs), a form of RNNs, for predicting missing parts in music \cite{liu2016predicting}. Pretrained neural networks have also been shown useful for music recommendation and auto-tagging, for instance by Liang et al.~\cite{liang2015content} and Van den Oord et al.~\cite{van2013deep}. Recently, Durand and Essid proposed a conditional random fields approach for downbeat detection, with features learned via deep architectures, in yet another example for combining graphical models with deep learning models \cite{durand2016downbeat}. Another deep generative approach that has been rising in prominence in recent years is that of Generative Adversarial Networks, or GANs, and indeed those too have been used in music AI tasks. As a recent example, Dong et al.\ proposed MuseGan, a symbolic-domain multi-track music synthesis framework trained on the Lakh dataset \cite{dong2018musegan}.

	
	Though somewhat beyond the scope of this paper, one of the most commonplace approaches for decomposing spectral data to individual components is that of \emph{matrix factorization methods}, which can be viewed as an unsupervised learning technique, and were mentioned when discussing music AI tasks, for instance the works of Panagakis et al., who presented a sparse multi-label linear embedding approach based on nonnegative tensor factorization and demonstrate its application to automatic tagging \cite{panagakisGDI}, or Kaiser et al., who used these factorization techniques to recover musical structure \cite{kaiser2010music}. To name a few more examples, Masuda et al.\ applied semi-supervised nonnegative matrix factorization for query phrase identification in polyphonic music \cite{masuda2014spotting}, while Sakaue et al.\ proposed a Bayesian nonnegative factorization approach for multipitch analysis cite{sakaue2012bayesian}. Liang et al.\ proposed a Beta process nonnegative factorization and show its potential usefulness in several tasks including blind source separation \cite{liang2013beta}, and subsequently Poisson matrix factorization for codebook-based music tagging \cite{liang2014codebook}.

Another large family of machine learning models that have seen frequent use in musical domains are \emph{decision trees}. To mention a few examples, Basili et al.\ applied decision trees for genre classification \cite{basili2004classification}. Lavner and Ruinskiy proposed a decision-tree based approach for fast segmentation of audio to music vs. speech \cite{lavner2009decision}. Herrera-Boyer and Peeters utilized a decision tree approach for instrument recognition \cite{herrera2003automatic}. West and Cox proposed a tree-based approach for learning optimal segmentations for genre classification \cite{west2005finding}.


As in other domains, the benefits of applying \emph{ensembles of classifiers} has not escaped the music informatics community. To mention a few examples, Tiemann et al.\ proposed an esnemble learning approach for music recommendation, generating many weak recommendations and combining them via learned decision templates \cite{tiemann2007ensemble}. Dupont and Ravet proposed a novel approach for instrument family classification using ensembles of t-SNE embeddings \cite{dupont2013improved}. Two particularly common ensemble approaches - boosting and random forests - have both been applied in music-related domains. Casagrande et al.\ used AdaBoost for frame-level audio feature extraction \cite{casagrande2005frame}. Turnbull et al.\ applied boosting for automatic boundary detection \cite{turnbull2007supervised}. Parker applied AdaBoost to improve a query-by-humming system \cite{parker2005applications}. Foucard et al.\ applied boosting for multiscale temporal fusion, later utilized for audio classification \cite{foucard2011multi}. In that paper, data from different timescales is merged through decision trees (serving as another example for the usage of this type of model in music tasks), which are then used as weak learners in an AdaBoost framework. The performance of their proposed system was tested on both instrument classification and song tag prediction, showing that their model was able to improve on prediction using features from only one timescale. Anglade et al.\ applied random forests to learn harmony rules, which were subsequently applied to improve genre classification \cite{anglade2009genre}.  
 

Lastly, it's worth mentioning that though it has not been applied as extensively as other techniques, evolutionary computation has also been used for various music tasks. For instance, Tokui and Iba proposed a system for interactive composition via evolutionary optimization (with human feedback serving as a fitness function) \cite{tokui2000music}. Biles adapted genetic algorithms for music improvization \cite{biles2007improvizing}, and as in Section \ref{chap8:gen}, Ramirez and Hazan employed genetic computation for expressive music performance \cite{ramirez2007inducing}.

While machine learning approaches may indeed be prevalent and ubiquitous in music (as in artificial intelligence research in general), other techniques have been applied as well. In the next subsection we will present two families of such methods: formal (or logic-based) approaches, and agent-based architectures.


			\subsection{Formal Methods}

While the learning-based approaches listed above are primarily data driven, many approaches have been employed for music tasks that are inherently rule-based and rely on formal reasoning. We consider this set of techniques as formal methods. 

Historically, one of the earliest approaches to the computational understanding of music involved linguistic analysis of music structure. Lehrdal and Jackendoff's seminal work on the generative theory of tonal music \cite{lerdahl1985generative} is one of the earliest examples for such an approach. Since then, many musicians and researchers have attempted to both analyze and generate music using the derivational structure of \emph{generative grammars} for music and other linguistic constructs \cite{rohrmeier2007generative, de2009modeling, mccormack1996grammar}. In a somewhat related work, Quick introduced the notion of chord spaces and applied concepts from Schenkerian analysis to define ``production rules'' for music generation \cite{quick2010generating}. 

As previously mentioned, Papadopoulos and Tzanetakis applied \emph{Markov Logic Networks} for modeling chord and key structure \cite{papadopoulos2012modeling}. Bergeron and Conklin proposed a structured pattern representation for polyphonic music that defined construction rules for hierarchical patterns, and utilize pattern matching techniques to extract such descriptions from symbolic data \cite{bergeron2008structured}. In another relevant example, Abdoli applied fuzzy logic to classify traditional Iranian music \cite{abdoli2011iranian}.
	
Lastly, though it has declined in fashion over the past 15 years, it is worth mentioning a sizable body of work	on music generation through constraint satisfaction techniques. This approach is typified by formulating music rules as constraints and using constraint solving techniques for music generation. For further details and examples, see Pachet and Roy's survey on harmonization with constraints \cite{pachet2001musical}.
	

			\subsection{Agent-Based Techniques}		

The definition of what exactly makes an ``agent'' is complicated and open for discussion, and it is outside the scope of this survey \cite{franklin1996agent}. For our purposes, we define an agent as an artificial system (either physical or, more commonly, implemented in software) that operates in an environment with which it interacts, and makes autonomous decisions. 

The vast majority of music-oriented robotics falls under this category. Robotic agents are autonomous systems which need to sense their environments, make decisions, and perform complex continuous control in order to achieve their goals. They may either need to play music alone, as in the work of Solis et al.\ on a robotic saxophone player \cite{solis2010development}, or with humans, as in the work of Hoffman et al.\ on a robotic marimba player \cite{hoffman2011interactive} and that of Peterson et al.\ on a robotic flute player \cite{petersen2010musical}, but their tasks still involve complex sensing and continuous control. Of course, not only physical robots serve as agents - autonomous accompaniment frameworks such as those proposed by Thom \cite{thom2001machine} and Raphael \cite{raphael2006demonstration} which we mentioned previously may certainly be considered autonomous agents. For a fairly recent survey of the state of the art in robotic musicianship, see \cite{bretan2016survey}.

Another family of approaches which we define as agent based are multiagent systems, where multiple autonomous, reactive components cooperate in order to perform a musical task. These approaches have been primarily utilized for music generation tasks. Examples include the swarm approach of Blackwell, previously mentioned in the context of music tasks. Blackwell modeled music through particle swarms which generate music through forces of attraction and repulsion \cite{blackwell2003swarm}. A somewhat similar approach can be seen in the more recent work of Albin et al., who utilized local properties in planar multi-robot configurations for decentralized real time algorithmic music generation \cite{albin2012musical}. 

Lastly, it is worth noting that some approaches have directly applied reinforcement learning, which is an agent-based learning paradigm, for various musical tasks. Cont et al.\ apply a reinforcement learning model for anticipatory musical style imitation \cite{cont2006anticipatory}. Wang et al.\ considered music recommendation as a multi-armed bandit problem, a concept closely related to the RL literature, with the explicit purpose of efficiently balancing exploration and exploitation when suggesting songs to listeners \cite{wang2014exploration}. And quite recently, 
Dorfer et al.\ framed score-following as a reinforcement learning task, a sensible approach given that changes in an agent estimation of its position in the score affect its expectation over future score position \cite{dorfer2018learning}. In that paper the authors also had the interesting insight that once the agent is trained, it does not need a reward function in order to generate predictions, an observation that would pave the road for other applications of reinforcement learning in similar situations.

To summarize, in this section we have reviewed the wide and varied range of artificial intelligence disciplines utilized in the context of music-related tasks. It is indeed apparent that nearly all major developments in artificial intelligence research have found their way to music applications and domains. In the next section we will address one of the primary challenges of music AI research - how do we evaluate algorithmic performance in music-related tasks?

\section{Evaluation Methods for Musical Intelligence Tasks}
\label{chap8:eval}

\begin{figure}[!htb]
\centering
\includegraphics[width=.8\linewidth]{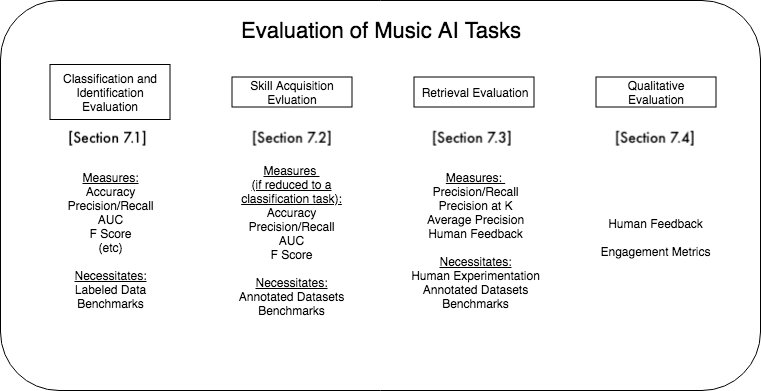}
\caption{Visual high-level overview of evaluation methods used in music AI research.} \label{chap8:fig_eval}
\end{figure}

Having delved into the vicissitudes of the music and AI literature, one should also consider the various evaluation metrics used in assessing success and failure in tackling the varied research questions previously mentioned. In this section we discuss the various approaches observed in the literature for evaluating performance on various musical tasks. Evaluation is often a challenge when it comes to the application of AI for music. Many musical tasks are inherently fuzzy and subjective, and on the face of it, any tasks that are aimed towards humans, be they music recommendation or affective performance, ultimately rely on human feedback as the most reliable (and perhaps the only) measure for success. An additional source of complication stems from the inherently sequential nature of music. In the case of image scene understanding, for instance, a person is able to perceive, recognize and annotate relatively quickly. Unlike visual data, music is experienced and processed sequentially in time, and often without being afforded the luxury of skipping information or ``speed auditing''. For these reasons, data from human participants is expensive to obtain, and various other methods have been employed in addition to it, depending on the task. We now briefly discuss such methods in this section. A visual illustration of the breakdown of evaluation method can be seen in Figure \ref{chap8:fig_eval}.

	\subsection{Evaluation of Classification Tasks}


One of the primary reasons why classification tasks have been popular in music informatics is its relative ease of evaluation. Given that a labeled dataset exists, evaluation can rely on the traditional evaluation metrics used in supervised learning, such as overall accuracy, AUC, F-scores etc \cite{murphy2012machine}. Some challenge may still lie in obtaining labeled examples. For certain tasks, such as classification by genre or composer, labels can easily be assigned automatically. For other tasks, such as guitar playing technique classification, getting label information is more difficult. In such cases, collecting hand-annotated data is a common solution  \cite{reboursiere2012left, su2014sparse}. Alternatively speculative labels may be inferred in some cases \cite{fu2011survey}. Another example of this kind of approach has been proposed recently by Sears et al., who described a data-driven method for the construction of harmonic corpora using chord onsets derived from the musical surface \cite{sears2018evaluating}.

Overall, for multiple tasks ranging from sentiment analysis and tagging in music to structure inference, preexisting hand-annotated datasets, such as the Mazurka project for performance analysis \cite{cook2007performance} or the various existing MIREX datasets \cite{downie2008music} serve as necessary benchmarks.

	\subsection{Evaluation of Skill Acquisition Tasks}


Skill acquisition (or music understanding) tasks, per our definition from Section \ref{chap8:tasks}, are generally more difficult than traditional classification, and as such tend to be more difficult to evaluate. For tasks such as music segmentation, structural analysis and motif identification, for instance, no trivial way to obtain ground truth information exists, and therefore most commonly researchers have relied on hand-annotated datasets for evaluation (as previously discussed in the context of classification tasks).

In certain contexts, in which the underlying skill is learned to facilitate a more complicated task, such as better genre classification, evaluation can be done directly on the final task. This observation holds for many of the aforementioned MIREX tasks, such as key detection and audio downbeat estimation, see the MIREX website for a most current list of tasks and benchmarks.\footnote{\url{https://www.music-ir.org/mirex/wiki/MIREX_HOME}}


In certain contexts, such as informative music performance, direct human evaluation has been applied, commonly in a comparative format (of the two performances, which one was more expressive?) \cite{lee2010crowdsourcing}.

One of the sources of difficulty evaluating skill acquisition tasks is the potential complexity of the ground truth data and metrics required in order to perform reliably. For instance, McLeod and Steedman note in a recent paper, in the context of evaluating polyphonic music transcription, that ``(i)t is less common to annotate this output with musical features such as voicing information, metrical structure, and harmonic information, though these are important aspects of a complete transcription''. In that paper they also propose a novel evaluation metric that combines different aspects of transcription that typically are evaluated separately, such as voice separation, pitch detection and metrical alignment. Despite such progress, the challenge of finding efficient and non-labor-intensive ways of evaluating musical skill acquisition tasks is not yet resolved.

	\subsection{Evaluation of Retrieval Tasks}


Like skill acquisition, retrieval tasks are nontrivial for evaluation. For example, they often rely on some notion of similarity among musical pieces, which is often a subjective and relative concept. Even when ground truth exists (for instance, in the form of playlists designed by humans \cite{mcfee2011natural}), deducing similarity or commonalities in taste is not immediate. 


For music recommendation systems, for instance, the best and most reliable evaluation method is through human experimentation, which is a difficult and time consuming process. Some researchers have gone around this by leveraging preexisting datasets as a surrogate for human preference \cite{mcfee2011natural}. Various different methods have been suggested to use limited existing data to impute speculative information regarding success or failure in the underlying task. For instance, in the context of playlist recommendation, it has been suggested that if a given person likes both artists A and B, then having songs by these two artists in the same playlist is considered a success \cite{chen2012playlist,weston2011multi}. In other tasks, such as mood analyis, particularly for retrieval purposes, given that certain songs by an artist have been labeled as ``moody'', assigning this label to other songs by that artist could be considered a success. These methods can be noisy and have obvious limitations (for instance, song style can vary considerably even for songs by the same artist). However, in a recent paper, Craw et al.\ have shown that preexisting datasets in combination with information extracted from social media can serve as a reasonable approximation for evaluating the effectiveness of different music recommenders, validating their approach via a human study \cite{craw2015music}.

	\subsection{Qualitative Evaluation}

Some music tasks, such as music generation, are very difficult to evaluate even with human feedback. For instance, the fact that 20 out of 100 human subjects liked or didn't like a song isn't in itself sufficient evidence for the quality of that song. For this reason, some researchers in the past have relied on qualitative evaluation by experts as a benchmark for performance. While such evaluation is foreign to the world of machine learning and artificial intelligence, it is in line with how culture in general is often evaluated. Another common approach aims for verisimilitude. In the case of style imitation, this approach has some legitimacy, though to the best of my knowledge very few if any recent algorithmic composition algorithms have been put to the test rigorously (i.e.\ having a large group of people try and differentiate between algorithmic compositions in the style of a given composer and pieces by that composer himself). If we were to speculate, I'd cautiously suggest that in most cases, even in light of recent, truly impressive advances in the field of generative music models (such as the work of Huang et al.~\cite{huang_transformer}), the differentiation between an actual composition by a renowned composer and an algorithmic one is either trivially easy (for experts in particular) or meaningless (for laymen, who would not be able to tell much less professional-sounding algorithmic approximation from actual human compositions). To conclude, despite much progress both in research and in analysis, the question of how to evaluate algorithmic composition in general remains an open problem.



\section{Summary \& Discussion: Open Problems}
\label{chap8:musint}
%
%
%
%
%
%
%
%
In this survey article we have reviewed an extremely large body of work involving both AI research and music-related tasks. We have proposed an overall taxonomy of music AI problems, breaking them down by the core nature of the task, by the input type, and by the algorithmic technique employed. We have then proceeded to map out the current state of the art, focusing on research from the past 20 years, relating a wide array of concrete exemplars to the proposed taxonomy.

This panoramic overview of music AI research reveals a dizzyingly complex picture, spanning disciplines and paradigms. On the one hand it feels as though almost any conceivable task has been attempted and any plausible technique has been employed. For some tasks, like key identification \cite{noland2006key} or beat detection \cite{durand2016downbeat}, the current levels of performance are high enough to allow for other tasks to rely on them as lower-level skills (for instance, key identification or beat and note extraction in the service of algorithmic accompaniment \cite{ramona2015capturing}, or score following \cite{otsuka2011incremental}). On the other hand, while the research community has been able to make significant strides on many music-related tasks spanning the gamut from extracting chords and notes to structure analysis to playlist recommendation to music synthesis, the more elusive goal of ``music understanding'' - as we proposed in Section \ref{chap:taxonomy} - is still largely unsolved. While we have been able to impart AI with the ability to identify many different building blocks necessary for music understanding, such as recognizing notes, chords, beats, motifs, sentiment (to some extent) and how these relate to more abstract things like listener preferences. But we have yet to teach AI to make sense of all these disparate sources of information; to ground their cultural and semiotic significance; understand the core characteristics of an individual's taste in music (and how it related to one's background, sense of identity etc); to know what a given chord means to a listener in a given setting; to understand what makes a piece by Telemann banal to modern ears and a piece by Bach a work of timeless genius; or to understand what people listen for when they listen to rock music vs. when they listen to a piano sonata by Beethoven.

In the next subsection we review the state of the art both with respect to specific tasks and from a higher-level perspective. In the subsequent subsection, we discuss the current gaps and limitations in the literature and what these gaps are indicative of, conceptually. Lastly, we consider possible ways forward in expanding the literature and bridging these gaps in the pursuit of more complete artificial musical intelligence.

\subsection{The State of the Art}

Examining the literature surveyed in this article reveals several insights regarding the current state of the art in applying machine learning approaches and tools in the context of music informatics. In this section we review the state of the art with respect to musical tasks, breaking it down along similar ones to those elucidated in Section \ref{chap8:tasks}.

\begin{itemize}

\item Over the past ten years, thanks to sustained research efforts and general advances in supervised machine learning techniques, performance on classification tasks such as instrument, genre and composer classification has been steadily growing. In a recent study, Oramas et al. reported AUC-ROC scores of up to 0.88-0.89 using audio information alone in a task of classifying music albums by genre\cite{oramas2018multimodal}, and Gomez et al. reported an F-score of 0.8 for Jazz solo instrument classification\cite{gomez2018jazz}. While this thread of research remains active and is expected to continue pushing the boundaries, it seems the community as a whole has gravitated towards more complex tasks which better fit the other categories of the task taxonomy - retrieval, skill acquisition and generation.

\item The dramatic increase in recommendation systems research and available online music consumption data has led to a boom in studies exploring music retrieval, recommendation, mood prediction and other user-facing tasks, as discussed at length in Section \ref{chap8:tasks}. Only recently, Schedl presented the LFM-1b Dataset, which contains $10^9$ listening events created by $12\cdot10^5$ users\cite{schedl2016lfm}, pushing the envelope even further with respect to the amount of data academic researchers can work with towards such tasks. Meanwhile, in industry, companies such as Spotify have over 200 million active users and 50 million tracks.\footnote{\url{https://newsroom.spotify.com/company-info/}} Despite this growth, the impression given by the literature is that progress in the quality of prediction for tasks such as music sentiment analysis and preference modeling is far from plateauing.

\item While improvements can always be made, existing approaches for fundamental music understanding tasks such as key and chord identification, beat extraction, audio transcription, score following, and even to some extent mood recognition, work well enough to provide serviceable performance as underlying components of more advanced tasks such as music recommendation and live accompaniment. This observation is supported by the increase in publications proposing such systems and their improved performance, requiring less direct human control or tuning.

\item In the past few years, harnessing the emergence of several discipline-altering advances in AI research such as deep neural network architectures, generative adversarial models, and attention mechanisms, huge strides have been made with the respect to AI-driven autonomous music generation, including Music Transformer\cite{huang_transformer}, MuseGan\cite{dong2018musegan} and MuseNet \cite{musenet}. While these advances are highly impressive, researchers \cite{chen2019effect} and musicians\footnote{\url{https://www.youtube.com/watch?v=xDqx14lZ_ls}} alike have commented on their existing limitation, highlighting the fact that AI-generated human-level music composition is still a challenge.

\end{itemize}

\subsection{Major Gaps in the Literature}

Examining the rich and varied work that has been carried out in pursuit of artificial musical intelligence, one may observe there has been an over-emphasis in the literature on isolating small, encapsulated tasks and focusing on them, without enough consideration of how different tasks connect to some end-goal vision of artificial intelligence for music. Despite their existence (as surveyed in this article, particularly under the category of agent-based techniques), there is a relative dearth of music AI \emph{systems}, entities that perform multiple music-related tasks over time, and connect music sensing, understanding and interaction. 

As a consequence of this gap, there has not been much work on music AI systems operating \emph{over time}. The challenge of end-to-end complex music reasoning systems is that they involve multiple facets of perception, abstraction and decision-making, not dissimilar from those of physical robotic or visual systems. While some progress has been made towards more robust and adaptive music AI capabilities, the conceptualization of music understanding as a process of sequential perception and decision-making over time is under-explored in the current literature.

Furthermore, there has not been much work on how such systems would practically interact with other agents and with humans and explicitly reason about their perceptions and intentions (for instance, in the context of joint human-agent music generation). More prosaically, the relative shortage of works which explicitly reason about people's \emph{perception} of music.

These gaps reflect not only a lack in music AI ``system engineering research'', i.e. the piecing together of different components towards an end-to-end functional architecture which is capable of sensing and acting in a closed loop fashion (though that is definitely part of the gap). They also indicate a conceptual lacuna with respect to modeling the \emph{implicit semantics} of music, understanding music hierarchically in a musicology-inspired fashioned to characterize in ways that go beyond statistical patterns and spectral subsequences what, on an abstract level, really makes two songs alike, or what characterizes one composer vs. the other.

Above all these challenges looms the fact that for many critical musical intelligence tasks, evaluation at scale is still an unresolved issue. As discussed in Section \ref{chap8:eval}, for any task complex enough such that labels cannot be automatically derived from the input, the curation of manually-annotated datasets is difficult and labor intensive. The difficulty of evaluation is substantially greater when it comes to music generation tasks, as no agreed upon metrics exist for ascertaining the quality of synthesized music, or for comparing pieces of synthesized music generated using different algorithms. 

In the next section, we propose a vision for music AI research which, in our opinion, would help put the community on a path forward towards meeting the challenges listed above.

\section{Directions Forward}

All in all, dramatic leaps forward have been made over the past decades in music informatics and the application of artificial intelligence techniques in musical tasks. However, as discussed in this section, the challenges remaining are substantial, and pose both technical and conceptual challenges. We believe that the conceptual challenges should be addressed irrespective of the many technical advances that are still being made by many researchers around the world. Here we propose a short, non-comprehensive list of concrete directions we believe offer the greatest opportunity for substantial progress:

\begin{itemize}

\item While isolated, well-scoped tasks are the building blocks with which progress is pursued, we believe it would be highly beneficial to the community to actively consider challenges of a bigger scale. Such challenge would introduce the need for end-to-end systems as well as a deeper conceptual understanding of what it means for AI to be musically intelligent. A good example for such a challenge would be a physical system that creates music while interacting with other musicians. Such a system should be required to actively sense what its collaborators are playing, reason about it abstractly, and generate audible sound in a closed-loop sense. Such a system would tie together challenges in music perception, music abstraction and modeling, prediction and decision-making, and would require anyone working on such a system to consciously consider how these various aspects of the problem really connect and inform one another. It is our hope that aiming towards such a goal would lead to substantial progress on each subtask individually, but more importantly, on our overall understanding of what synthetic music competency means.

\item While there has been huge progress in the creation of large-scale, meaningfully annotated music datasets for AI research, there is still no ``ImageNet\cite{deng2009imagenet} equivalent'' for music. We believe a benchmark of such nature - a rich, audio-level dataset with complex annotations on a massively grand scale - would lead to considerable progress and would not only push the field forward but also serve as a consistent shared baseline across algorithms and platforms, even beyond music informatics. More importantly, if the goal of algorithmic, AI-driven music synthesis is truly a tent-pole for music AI research, we must strive for some shared notion of a metric or evaluative procedure for comparing the outputs of such synthesized pieces of music, a measure which goes beyond collective impressions. A possible approach towards addressing the issue of evaluating AI-generated music could be a formal competition, with some credentialed experts as referees and predefined criteria. Such an expert panel approach could be complemented by a more traditional crowdsourced approach. Together, these two formats of evaluation could provide us with a clearer picture of how the music establishment as well as the general public view these generated pieces comparatively.

\item Lastly, we believe there is a great deal to be gained in bridging the gap between music AI and cognitive research. Music is an innate form of human communication. How we perceive music and reason about it should be made a more integral aspect of music AI research. First, because ultimately any music AI tool would need to interact with human perception in some way. Second, because leveraging a better understanding of human music cognition could inform better music AI algorithms. And lastly, because in the process we might also learn something profound about our own music cognition, and how it is related to other facets of our perception and reasoning.

\end{itemize}

\subsection{Concluding Remarks}

If we envision a future where intelligent artificial agents interact with humans, we would like to make this interaction as natural as possible. We would therefore like to give AI the ability to understand and communicate within cultural settings, by correctly modeling and interpreting human perception and responses. Such progress would have many real world practical benefits, from recommender systems and business intelligence to negotiations and personalized human-computer interaction. 

Beyond its practical usefulness, having AI tackle complex cultural domains, which require advanced cognitive skills, would signify a meaningful breakthrough for AI research in general. The dissertation research of the first author of this survey was largely motivated by the desire to address the gaps discussed in the previous section, particularly on work towards the goal of learning social agents in the music domain\cite{Liebman2020}. However, the progress made in one dissertation only highlights how much challenging work is left to be pursued. We believe this work presents incredible opportunities for musical intelligence, and for artificial intelligence as a whole.

\pagebreak
\section*{Musical Terms}
\label{app:glossary}

\begin{table}[h!]
\begin{tabular}{ll}
{\bf term}                 & {\bf meaning}                                                                                        \\
beat        & basic unit of time        \\
chord       & concurrent set of notes   \\
interval    & a step (either sequential or concurrent) between notes  \\ 
loudness    & amplitude of audible sound      \\
major chord & a chord based on a major third interval     \\
minor chord & a chord based on a minor third interval    \\
note        & sustained sound with a specific pitch   \\
pitch       & the perceived base frequency of a note     \\
playlist    & ordered sequence of songs \\
tempo       & speed or pace of a given music       \\
timbre      & perceived sound quality of a given note                       
\end{tabular}
\end{table}
\newpage

\section*{Bibliography}
\bibliography{refs_prop}
\bibliographystyle{elsarticle-num}

\end{document}